\documentclass{aastex631}
\usepackage{CJKutf8}

\shorttitle{Reinvestigation of FRBs HG and Rate}
\shortauthors{Ma et al.}
\graphicspath{{./}{figures/}}

\begin{document}

\title{Reinvestigation of Fast Radio Burst Host Galaxy and Event Rate Density}

\author[0009-0002-7690-7381]{W. Q. Ma (\begin{CJK*}{UTF8}{gbsn}马文琦\end{CJK*})}
\affiliation{Xinjiang Astronomical Observatory, Chinese Academy of Sciences, 150 Science 1-Street, Urumqi, Xinjiang 830011, China}
\affiliation{University of Chinese Academy of Sciences, No. 19 Yuquan Road, Beijing 100049, China}
\affiliation{Key Laboratory of Radio Astronomy, Chinese Academy of Sciences, West Beijing Road, Nanjing 210008, China}

\author[0000-0002-0138-3360]{Z. F. Gao (\begin{CJK*}{UTF8}{gbsn}高志福\end{CJK*})}
\altaffiliation{zhifugao@xao.ac.cn}
\affiliation{Xinjiang Astronomical Observatory, Chinese Academy of Sciences, 150 Science 1-Street, Urumqi, Xinjiang 830011, China}
\affiliation{University of Chinese Academy of Sciences, No. 19 Yuquan Road, Beijing 100049, China}


\author[0000-0003-0325-6426]{B. P. Li (\begin{CJK*}{UTF8}{gbsn}李彪鹏\end{CJK*})}
\affiliation{Xinjiang Astronomical Observatory, Chinese Academy of Sciences, 150 Science 1-Street, Urumqi, Xinjiang 830011, China}
\affiliation{University of Chinese Academy of Sciences, No. 19 Yuquan Road, Beijing 100049, China}
\affiliation{Key Laboratory of Radio Astronomy, Chinese Academy of Sciences, West Beijing Road, Nanjing 210008, China}

\author[0000-0001-6651-7799]{C. H. Niu (\begin{CJK*}{UTF8}{gbsn}牛晨辉\end{CJK*})}
\affiliation{Institute of Astrophysics, Central China Normal University, Wuhan 430079, China}

\author[0000-0002-4997-045X]{J. M. Yao (\begin{CJK*}{UTF8}{gbsn}姚菊枚\end{CJK*})}
\affiliation{Xinjiang Astronomical Observatory, Chinese Academy of Sciences, 150 Science 1-Street, Urumqi, Xinjiang 830011, China}
\affiliation{University of Chinese Academy of Sciences, No. 19 Yuquan Road, Beijing 100049, China}

\author[0000-0003-4157-7714]{F. Y. Wang (\begin{CJK*}{UTF8}{gbsn}王发印\end{CJK*})}
\altaffiliation{fayinwang@nju.edu.cn}
\affiliation{School of Astronomy and Space Science, Nanjing University, Nanjing 210093, China}
\affiliation{Key Laboratory of Modern Astronomy and Astrophysics (Nanjing University), Ministry of Education, Nanjing 210093, China}



\begin{abstract}

Fast radio bursts (FRBs) are radio signals that last milliseconds. They originate from cosmological distances and have relatively high dispersion measures (DMs), making them being excellent distance indicators. However, the origins of the FRB remain to be resolved. With its wide field of view and excellent sensitivity, CHIME/FRB has discovered more than half of all known FRBs. As more and more FRBs are located within or connected with their host galaxies, the study of FRB progenitors is becoming more important. In this work, we collect the currently available information related to the host galaxies of FRBs, and the Markov Chain Monte Carlo analysis about limited localized samples reveals no significant difference in the median of $\mathrm{DM_{host}}$ between repeaters and non-repeaters. After examining and selecting CHIME/FRB samples, we estimate the local volumetric rate density of repeaters and non-repeaters, accounting for different $\mathrm{DM_{host}}$ contributions, and compare with rates of predicted origin models and transient events. Our results indicate that $\mathrm{DM_{host}}$ significantly affects volumetric rates and offer insights into the origin mechanisms of FRB populations.

\end{abstract}


\section{Introduction}
\label{sec:intro}

Fast radio bursts (FRBs) are bright radio bursts that originated from cosmological distances, with pulse durations lasting milliseconds, and have been detected in a wide band range (between $110~\text{MHz}$ and $8~\text{GHz}$) \citep{2018ApJ...863....2G, 2021Natur.596..505P}. Research on FRBs origins, mechanisms and applications is currently one of the hot topics in astrophysics \citep{2019A&ARv..27....4P,2022A&ARv..30....2P,  2021SCPMA..6449501X, 2023RvMP...95c5005Z}.

In 2007, \citet{2007Sci...318..777L} discovered the first FRB signal in historical observation data from the Parkes telescope. The dispersion measure (DM) far exceeding the estimates for the Milky Way indicating its origin at cosmological distances. FRB 20121102A was initially detected by the Arecibo Telescope, also suggesting its extragalactic origin \citep{2014ApJ...790..101S}. The follow-up observation of FRB 20121102A found additional bursts from the same sky position, identifying it as the first repeater \citep{2016Natur.531..202S}. Localization studies of repeater FRB 20121102A confirmed that it originates from a dwarf galaxy at $z = 0.19$ \citep{2017ApJ...834L...7T}, further establishing the cosmological origin of FRBs.

On April 28, 2020, the CHIME/FRB and Survey for Transient Astronomical Radio Emission 2 both detected a signal similar to an FRB \citep{2020Natur.587...59B, 2020Natur.587...54C}, revealing the extremely high isotropic equivalent energy ($E_\mathrm{iso}\sim (10^{34}-10^{35})~\mathrm{erg}$). Swift reported multiple bursts from the galactic magnetar SGR 1935+2154 before, which was identical to the position of FRB 200428. Surprisingly, X-ray bursts from SGR 1935+2154 associated with FRB 20200428 have been confirmed \citep{2020Natur.587...54C, 2020ApJ...898L..29M, 2021NatAs...5..378L, 2021NatAs...5..372R, 2021NatAs...5..401T}. Considering magnetars (highly magnetized neutron stars) can occasionally produce enormous bursts and flares of X-rays and gamma rays, magnetars have become a possible progenitor model \citep{2021ApJ...907L..31B}. The individual pulse emission from radio pulsars, characterized by short and intense bursts of radio waves \citep{2016Ap&SS.361..261W, 2016A&A...592A.127W, 2022ApJ...929...71W}, shares striking similarities with FRBs, suggesting that both may originate from extreme astrophysical environments involving intense magnetic fields and particle acceleration. However, it remains uncertain whether FRBs from cosmological distances could also originate from magnetars.

Among the currently known FRBs, non-repeaters account for a significant proportion, with less than a hundred FRBs bursting repeatedly. Consequently, it is believed that repeaters and non-repeaters have different origins. For repeaters, monitoring their bursts is a rather time-consuming task because their activity is rare and often unpredictable. Long-term monitoring has shown that some bursts exhibit potential periodic activity windows \citep{2020Natur.582..351C, 2020MNRAS.495.3551R}, but these results still require corresponding physical models for further explanation. Information on  burst energy and the isotropic equivalent luminosity of FRB 20121102A can be obtained by combining burst samples with redshift. Analysis of localized FRBs and their host galaxies has revealed that FRBs may originate from a variety of host galaxies \citep{2021ApJ...910L..18B, 2022Natur.606..873N}. Therefore, more localization samples are needed to study the differences and connections between repeaters and non-repeaters, as well as their host galaxies and surrounding environments.

Achieving arcseconds (even sub-arcseconds) localization for FRBs is quite difficult, limiting the study of FRB host galaxies. Among the localized FRBs, noteworthy are not only those with high host galaxy DM contributions but also those at low redshifts. For instance, FRB 20200428 indicates that magnetars are one of the origins of FRBs. Another repeater, FRB 20200120E, is also worth discussing, it is firstly localized near the Milky Way's neighboring galaxy M81 by \citet{2021ApJ...910L..18B}, with the estimated redshift of less than 0.03. However, it was found that FRB 20200120E was offset by about 20 kpc from the center of the M81 galaxy. Later, \citet{2022Natur.602..585K} further localized it to a globular cluster associated with M81. Analysis of FRB 20200120E suggests that the DM contribution from the host galaxy M81 halo is about $(15-50)~\text{pc~cm}^{-3}$, while the contribution from the galactic disk and the globular cluster itself is negligible, thus constraining the DM contribution from the Milky Way halo to be less than $(32-42)~\text{pc~cm}^{-3}$ \citep{2022Natur.602..585K}. This result is consistent with the $\mathrm{DM_{halo}}$ and $\mathrm{DM_{host}}$ inferred using the YMW16 and YT20 models. Moreover, some localized FRBs are associated with the star-forming regions of galaxies, implying a connection to the channel of forming young magnetars through core-collapse supernovae (CCSNe), as the event rates of accretion-induced collapse and merger-induced collapse are lower \citep{2019ApJ...886..110M}.

The Canadian Hydrogen Intensity Mapping Experiment (CHIME) is a drift-scan radio telescope located at the Dominion Radio Astrophysical Observatory, Canada. It consists of four north–south oriented parallel cylindrical reflectors, each one is $100~\mathrm{m}\times20~\mathrm{m}$ and outfitted with a 256-element dual-polarization linear feed array \citep{2018ApJ...863...48C, 2022ApJS..261...29C}. CHIME, operating in the 400-800 MHz band, features a large field of view and high sensitivity, making it one of the best telescopes for detecting FRBs \citep{2019Natur.566..230C, 2019Natur.566..235C, 2019ApJ...885L..24C, 2021ApJS..257...59C, 2023ApJ...947...83C, 2020ApJ...891L...6F}.

DM is an excellent distance indicator. Considering the challenge of pinpointing the location of FRBs, DM provides a feasible way to estimate distance. Clarifying the contributions of different DM components of FRBs helps us study various properties of these phenomena. However, due to the unclear contribution of the host galaxy's DM, a fixed value or uniform distribution is assumed in some studies of event rate \citep{2019NatAs...3..928R, 2020ApJ...893...44Z}. The event rate of FRBs is a crucial pathway to study their origins. Several papers that discuss the event rate for different luminosity/energy ranges have been published, and the redshift evolution of the FRB has been discussed, providing an important clue to constrain the FRB origins \citep{2018MNRAS.481.2320L, 2020MNRAS.494..665L, 2022MNRAS.510L..18J, 2022MNRAS.511.1961H, 2023ApJ...944..105S, 2024ApJ...973L..54C}. However, due to the small number of FRBs observed in the high redshift region, the luminosity/energy function and redshift evolution remain uncertain. Therefore, comparing the local volumetric rate with other progenitor models is a more viable approach.

In this study, we first analyze the localized sample of FRBs and discuss the properties of host galaxies of both repeaters and non-repeaters. Based on the assumption that repeaters and non-repeaters have different environments, we utilize FRB sources discovered by CHIME/FRB to estimate the volumetric rates of different FRB populations under various host galaxy DM contributions. The results will provide some reference for studying the origin mechanisms of potential FRB populations. This paper is organized as follows. In Section \ref{sec:theory}, we introduce the theories regarding DM contributions from FRBs and methods for calculating event rates. In Section \ref{sec:results}, we present the results of event rate estimations for different $\mathrm{DM_{host}}$ contributions and compare them with FRB progenitors. Our conclusions and discussion are shown in Section \ref{sec:discussion}.

\section{Theories and Methods}
\label{sec:theory}

During the propagation of radio signals, the interaction of different frequencies of electromagnetic waves with free electrons in the medium causes differences in the arrival times of low-frequency and high-frequency signals. Thus, the time delay of a pulse between the highest and lowest observational radio frequencies is quantified as \citep{2019A&ARv..27....4P}

\begin{equation}
\Delta t = \frac{e^2}{2\pi m_e c}(\nu_\mathrm{lo}^{-2}-\nu_\mathrm{hi}^{-2})~\mathrm{DM}\approx 4.15(\nu_\mathrm{lo}^{-2}-\nu_\mathrm{hi}^{-2})~\mathrm{DM~ms},\label{eq:01}
\end{equation}

\noindent where $m_e$ is the electron rest mass, $c$ is the speed of light in vacuum. The second approximation is valid only when $\nu_\mathrm{lo}$ and $\nu_\mathrm{hi}$ are in units of $\mathrm{GHz}$. The DM in units of $\text{pc~cm}^{-3}$ in Equation (\ref{eq:01}) is the integral of the electron density along the propagation path and calculated as

\begin{equation}
\mathrm{DM} = \int_{0}^{d} n_e(l)\mathrm{d}l,\label{eq:02}
\end{equation}

\noindent where $n_e$, $l$, and $d$ represent the electron number density, path length, and distance of the source, respectively. Due to the cosmological origin, the DM from an FRB can be defined as \citep{2020ApJ...893...44Z, 2023RvMP...95c5005Z}

\begin{equation}
\mathrm{DM_{FRB}}=\mathrm{DM_{MW}}+\mathrm{DM_{halo}}+\mathrm{DM_{IGM}}+\frac{\mathrm{DM_{host}}+\mathrm{DM_{source}}}{1+z}.\label{eq:03}
\end{equation}

\noindent Here, $\mathrm{DM_{MW}}$ and $\mathrm{DM_{halo}}$ are the contributions from the Milky Way interstellar medium (ISM) and outer halo, respectively, $\mathrm{DM_{IGM}}$ is the contribution from the intergalactic medium (IGM), while $\mathrm{DM_{host}}$ and $\mathrm{DM_{source}}$ are the contributions from the host galaxy and the surrounding medium of source, respectively. The factor $1 + z$ accounts for the cosmological time dilation due to redshift, allowing us to convert these two part contributions to the frame of sources. We define extragalactic dispersion measure $\mathrm{DM_{ex}}$ as

\begin{equation}
\mathrm{DM_{ex}} = \mathrm{DM_{FRB}}-\mathrm{DM_{MW}}-\mathrm{DM_{halo}} = \mathrm{DM_{IGM}}+\frac{\mathrm{DM_{host}}+\mathrm{DM_{source}}}{1+z}.\label{eq:04}
\end{equation}

The DM of FRBs is a typical distance indicator. We will show how to handle the contributions of each part in Sections \ref{milky way} - \ref{host}.

\subsection{Contribution from Milky Way and Halo}
\label{milky way}
The ISM in the disk of the Milky Way and the extended Galactic halo contributes to the $\mathrm{DM_{MW}}$ and $\mathrm{DM_{halo}}$. If we know the electron density along the line of sight, we can determine the DM contribution from the Milky Way, as well as the distance to the pulsar, which requires the galactic electron density model (GEDM).

The DM contribution from the Milky Way can be estimated using the NE2001 \citep{2002astro.ph..7156C} or YMW16 \citep{2017ApJ...835...29Y} model, which incorporate distance and scattering measurements from pulsars as well as objects such as globular clusters and active galactic nuclei. Here, we calculate the $\mathrm{DM_{MW}}$ of FRBs using the YMW16 model. Notice that using the NE2001 model for estimation does not affect subsequent research results.

We use the YT20 model to calculate the contribution from halo. The YT20 model is a dark matter halo model of the Milky Way proposed by \citet{2020ApJ...888..105Y}. By considering recent diffuse X-ray observations, they constructed a new extended hot gas halo model encompassing both a disk-shaped and a spherical halo, and provided corresponding analytical formulas, making it easier to estimate the DM contribution from the Milky Way's halo in the direction of extragalactic sources. Their results show that the disk component is very important for interpreting the currently observed direction dependence of the electromagnetic field, estimating $\mathrm{DM_{halo}}$ to be between $30$ and $245~\text{pc~cm}^{-3}$ over the entire sky, with an average of $43~\text{pc~cm}^{-3}$.

We utilize the PyGEDM software package to calculate the independent contributions. PyGEDM is a Python package that provides a unified interface for GEDM (NE2001 or YMW16) and halo model (YT20) \citep{2021PASA...38...38P}. The PyGEDM code is open source and freely available online\footnote{\url{https://github.com/FRBs/pygedm}}, and it also provides a web app to convert between distance and DM for the NE2001 and YMW16 models \footnote{\url{http://apps.datacentral.org.au/pygedm/}}. By using the PyGEDM package, the DM from the Milky Way and halo for a specific model in any line of sight can be conveniently calculated and integrated into projects.

\subsection{Contribution from the IGM}
\label{igm}

Since we estimate the distance of FRBs directly through the DM from the IGM, studying the distribution of $\mathrm{DM_{IGM}}$ is very important, as it relates to the distribution of matter in the universe.

The dispersion measure distribution of $\mathrm{DM_{IGM}}$ can be fitted as \citep{2020Natur.581..391M, 2021ApJ...906...49Z}

\begin{equation}
P(\Delta)=A\Delta^{-\beta}e^{-\frac{(\Delta^{-\alpha}-C_0)^2}{2\alpha^2\sigma_\mathrm{DM}^2}},\quad\Delta>0,\label{eq:05}
\end{equation}

\noindent where $\Delta=\frac{\mathrm{DM_{IGM}}}{\langle\mathrm{DM_{IGM}}\rangle}$, $A$ is a normalization parameter, the parameters $\alpha=\beta=3$ are related to the slope and density profile of the gas in halos, $\sigma_\mathrm{DM}$ represents the standard deviation of dispersion, $C_0$ is a free parameter. All parameter values in Equation (\ref{eq:05}) are adopted from the results in \citet{2021ApJ...906...49Z}, who simulated all parameters using the IllustrisTNG project. The average IGM contribution at redshift $z$, ${\langle\mathrm{DM_{IGM}}\rangle}$, is calculated as follows \citep{2018ApJ...852L..11S, 2021ApJ...906...49Z}:

\begin{equation}
\langle\mathrm{DM_{IGM}}\rangle=\frac{3c\Omega_bH_0}{8\pi G m_p}\int_{0}^{z}\frac{(1+z^\prime)f_\mathrm{IGM}(z^\prime)f_e(z^\prime)}{\sqrt{{(1+z^\prime)}^3\Omega_m+\Omega_\Lambda}}\mathrm{d}z^\prime,\label{eq:06}
\end{equation}

\noindent where $H_0$ is the Hubble constant and $m_p$ is the mass of proton. $f_\mathrm{{IGM}}$ represents the fraction of baryon mass in the IGM \citep{2012ApJ...759...23S}. $f_e=Y_\mathrm{H} X_{e,\mathrm{H}}(z)+\frac{1}{2}Y_\mathrm{He} X_{e,\mathrm{He}}(z)$, $Y_\mathrm{H} = 3/4$ and $Y_\mathrm{He} = 1/4$ are the mass fractions of hydrogen and helium, respectively. And $X_{e,\mathrm{H}}(z)$ and $X_{e,\mathrm{He}}(z)$ are the ionization fractions of intergalactic hydrogen and helium. The cosmological parameters are chosen from the Planck18 results \citep{2020A&A...641A...6P}: $H_0=67.74~\mathrm{km~s^{-1}~{Mpc}^{-1}}$, $\Omega_m=0.3111$ and $\Omega_\Lambda=0.6889$.

\begin{figure}[htb!]
    \centering
    \includegraphics[width=0.45\textwidth]{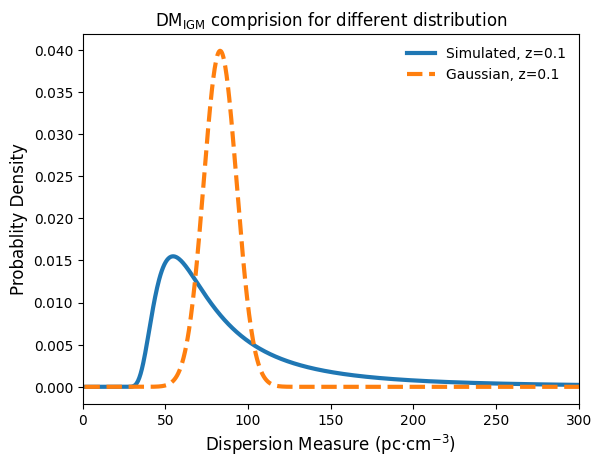}
    \hspace{0.1in}
    \includegraphics[width=0.45\textwidth]{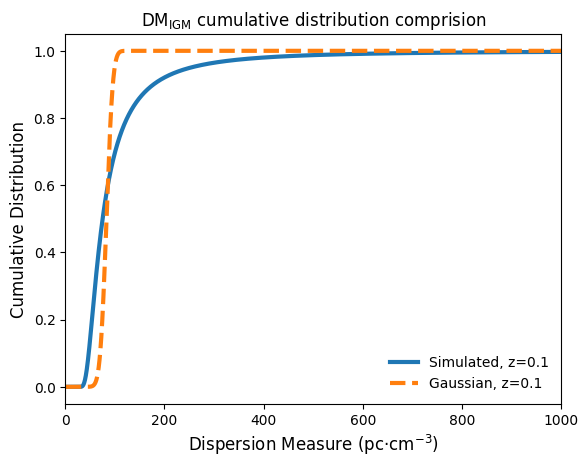}
    \caption{Left panel: probability density comparison for two $\mathrm{DM_{IGM}}$ distributions. Right panel: cumulative distribution comparison for two $\mathrm{DM_{IGM}}$ distributions.}
    \label{fig:01}
\end{figure}

If we assume that the $\mathrm{DM_{IGM}}$ distribution in a specific redshift follows a normal distribution with a constant error \citep{2019NatAs...3..928R, 2020ApJ...893...44Z}, it will result in a significantly different distance estimation. Figure \ref{fig:01} shows the probability density functions for two scenarios (i.e. normal and simulated distribution) and a comparison of the cumulative distribution functions. The simulated distribution suggests that $\mathrm{DM_{IGM}}$ has a different median and is more dispersed than the normal distribution at the same redshift. Therefore, this results in a wider distribution and a tailing effect, causing differences in distance estimation.

\subsection{Contribution from Host Galaxy and Source}
\label{host}

The host galaxies of FRBs exhibit a series of properties in terms of offset, stellar masses, star formation rates (SFRs), metallicities, etc. The host galaxy can also provide messages about the progenitors. For example, repeaters and non-repeaters can be born from diverse progenitors, resulting in varying environments \citep{2021ApJ...907..111Z, 2022Natur.606..873N}. The first repeater, FRB 20121102A, is believed to be closely related to a persistent radio source \citep{2017ApJ...834L...7T}. FRB 20190520B, discovered by the Five-hundred-meter Aperture Spherical radio Telescope, was localized to a dwarf galaxy at $z\sim0.241$ using Very Large Array (VLA) observations, inferring a very large $\mathrm{DM_{host}}$ contribution \citep{2022Natur.606..873N}. Both repeaters are related to the complex electromagnetic environment near the sources and also suggest that young, active repeater sources may have complex origins.

A large sample of host galaxies will provide further insights into FRB progenitors, and could reveal conclusive evidence that the FRB progenitor population evolves with cosmological time (stellar mass or SFR). Since the DM contributed by the host galaxy is coupled with that from the IGM, it is challenging to distinguish the contribution from the host galaxy. Currently, there are more than 60 FRBs that have been localized or associated with potential host galaxy candidates. Here, we summarize the localized FRBs and the host galaxy properties. The detailed data are listed in the Appendix \ref{appendix:A}.

From the Appendix \ref{appendix:A}, we find that almost all localized FRBs have redshift within the range of $z<1$. The diversity in host galaxy morphology, stellar mass, and SFR exhibits different characteristics, indicating that the birth environments of FRBs are complex and variable. For instance, there are only four FRBs localized to dwarf galaxies. The hosts of FRB 20121102A and FRB 20190520B, however, exhibit significantly higher SFRs than those of the other two FRB host galaxies \citep{2017ApJ...834L...7T, 2022Natur.606..873N}. A lower SFR may not support models evolving along the SFR. Additionally, the host galaxies of several FRBs (e.g. FRB 20121102A) also show the characteristics of low metallicity, properties closely linked to long gamma-ray bursts (LGRBs) and super-luminous supernovae events \citep{2017ApJ...834L...7T}.

\begin{figure}[htbp]
    \includegraphics[width=0.45\textwidth]{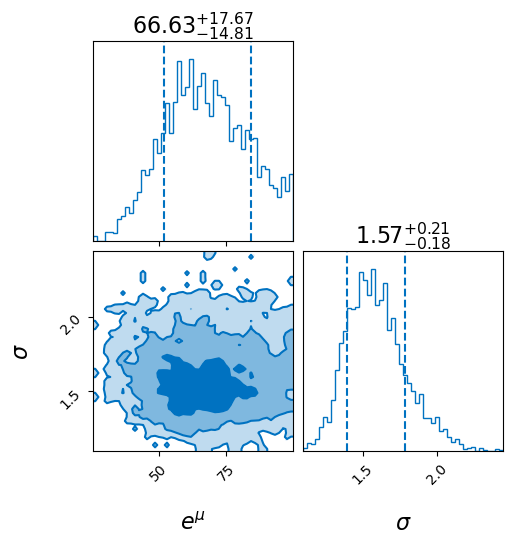}
    \hspace{0.1in}
    \includegraphics[width=0.45\textwidth]{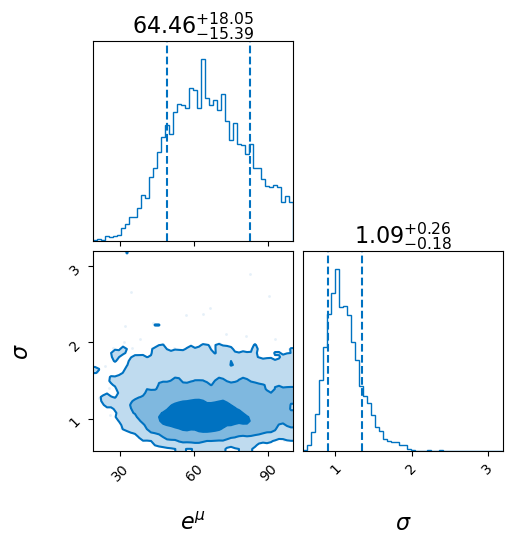}
    \caption{MCMC results for localized non-repeaters (left, 35 in total) and repeaters (right, 14 in total).}
    \label{fig:02}
\end{figure}

Formally, the log-normal distribution is used to describe the distribution of $\mathrm{DM_{host}}$\citep{2020ApJ...900..170Z}:

\begin{equation}
P(\mathrm{DM_{host}})=\frac{1}{\sqrt{2\pi}\sigma_\mathrm{host}\mathrm{DM_{host}}}e^{-\frac{(\mathrm{ln~DM_{host}-\mu_{host})}^2}{2\sigma_\mathrm{host}^2}},\label{eq:07}
\end{equation}

\noindent where $\mu_\mathrm{host}$ represents the median, and $\sigma_\mathrm{host}$ is the logarithmic width parameter. Note that we do not consider the evolution of ionized gas in galaxies and galaxy clusters with the cosmic star formation history and its impact on the evolution of DM with redshift. We utilize the Python package bilby \footnote{\url{https://github.com/bilby-dev/bilby}} to perform Markov Chain Monte Carlo (MCMC) analysis of $\mathrm{DM_{host}}$ for non-repeaters and repeaters \citep{bilby_paper}. We only use sources with positive inferred $\mathrm{DM_{host}}$ and the specific results are shown in Figure \ref{fig:02}. Based on the current localized samples, we find no significant difference between repeaters and non-repeaters, suggesting that they originate from similar environments. It is also possible that many non-repeaters are actually repeaters but are too faint to be detected.

Observing the direct environment of an FRB requires precise localization, preferably to subarcsecond precision so as to identify the local environment within the host galaxy. As more and more FRBs are localized to their host galaxies, our understanding of the host galaxies has improved. Meanwhile, the DM contribution from the source has become increasingly significant in the progenitor models. Although the rotational measure (RM) from the source is easier to identify, high RM values indicate a highly magnetized and ionized local environment. A few FRBs have now been identified with significant RM excess, suggesting that substantial magneto-ionic material local to the source is relatively common for repeaters, even for some non-repeaters \citep{2022Natur.609..685X, 2023Sci...380..599A}.

Unlike pulsars in the Milky Way, where DM variations are primarily caused by the source, such variations are expected to be negligible for distant FRBs. However, considering that repeaters may originate in dense, dynamic environments, such as expanding supernova remnants, their DM might show significant changes \citep{2017ApJ...847...22Y, 2018ApJ...861..150P}.

Long-term monitoring of variations in DM, RM, and scattering of FRBs can place constraints on the local magneto-ionic material in their progenitor environment. Some studies point out that if FRBs originate from CCSNe (Type I supernovae (SN Ia), they would generate a considerable $\mathrm{DM_{source}}$ contribution, whereas the $\mathrm{DM_{source}}$ contribution from binary mergers would be much smaller \citep{2023RvMP...95c5005Z}. Observations at low radio frequencies ($\sim100~\mathrm{GHz}$) can directly limit the density of the local environment by constraining absorption in the local ionized medium \citep{2021Natur.596..505P}. Since the progenitors of FRBs are still uncertain, we do not consider its impact on our research and combine this contribution with $\mathrm{DM_{host}}$ for the following analysis.

\subsection{Volumetric Event Rate Density Estimation}

Studying the volumetric event rate density is crucial for understanding the origins of FRBs. If we assume the occurrence of an FRB follows a Poisson process, meaning that observed FRBs are independently and identically distributed both spatially and temporally, we can regard their number as following a Poisson distribution.

Based on the methods proposed by \citet{2019NatAs...3..928R} and \citet{2020ApJ...893...44Z}, we estimate the volumetric event rate density by

\begin{equation}
R(\ge F_\mathrm{lim})=\frac{N/\epsilon}{V_\mathrm{limit}\ast\Omega_\mathrm{sky}\ast t_\mathrm{obs}\ast f_b}\ \mathrm{Gpc^{-3}~yr^{-1}},\label{eq:08}
\end{equation}

\noindent where $F_\mathrm{lim}$ represent the observed fluence lower limit of CHIME/FRB, $V_\mathrm{limit}=\frac{4}{3}\pi{d_\mathrm{limit}^3}$ is the comoving volume that covers FRBs, $N$ is the number of FRBs in this volume, $\Omega_\mathrm{sky}=\frac{250~\mathrm{deg}^2}{41252.96~\mathrm{deg}^2}\approx0.006$ is the sky coverage of CHIME \citep{2018ApJ...863...48C}, $t_\mathrm{obs}$ is the total observation time, $f_b=0.1$ is the beaming factor \citep{2017ApJ...843...84N}, and $\epsilon = 39638/84697 \approx 0.468$ is the detection efficiency derived from the injection test \citep{2021ApJS..257...59C}. The errors at different confidence levels are calculated using the approximate formula given by \citet{1986ApJ...303..336G}.

To estimate the volumetric rate in a certain space, we need to determine the size of $V_\mathrm{lim}$ (i.e. the maximum of comoving distance) from Equation (\ref{eq:08}). For an FRB with an extragalactic dispersion measure of $\mathrm{DM_{ex}}$, we consider the probability that an FRB with inferred $\mathrm{DM_{ex}}$ actually exists in space with volume $V_\mathrm{lim}$ as $P(\le d_\mathrm{limit}|\mathrm{DM_{ex}})$, which is the probability it lies within a cosmological distance $d_\mathrm{limit}$. We filter FRBs by selecting $\mathrm{DM_{ex}}<\mathrm{DM_{IGM}}(z)$ to determine the size of the cosmic volume. Specifically, we set the FRB with the highest $\mathrm{DM_{ex}}$ in the selected samples to satisfy $P(\le d_\mathrm{limit}|\mathrm{DM_{ex}})\approx0.95$, and ensure that all other FRBs satisfy $P(\le d_\mathrm{limit}|\mathrm{DM_{ex}})\approx1$. Due to the relation between $\mathrm{DM_{IGM}}$ and $\mathrm{DM_{host}}$ and the importance of $\mathrm{DM_{host}}$ in distance estimation, we cannot disregard this contribution. Based on the analysis from localized FRBs, we will consider different range of $\mathrm{DM_{host}}$ for the volumetric rate estimation.

\subsection{Sample Selection}

We conduct the analysis only using sources observed by CHIME/FRB. We only select non-repeater samples collected in the CHIME/FRB Catalog 1 \citep{2021ApJS..257...59C}, which contains 474 non-repeaters and 18 repeaters. For repeaters, we use data collected by the \textit{Blinkverse} database. \textit{Blinkverse}\footnote{\url{https://blinkverse.zero2x.org/overview}} is an FRB database established and maintained by Zhejiang Lab, China \citep{2023Univ....9..330X}. All repeaters were discovered before 2023 December 31 and collected from several CHIME observations \citep{2019ApJ...885L..24C, 2023ApJ...947...83C, 2020ApJ...891L...6F}. We select samples following these data selection criterion:

\begin{enumerate}
    \item Events with $\mathrm{DM_{obs}}>1.5~max(\mathrm{DM_{NE2001}}, \mathrm{DM_{YMW16}})$.
    \item Events with $\mathrm{DM_{obs}}>100~\mathrm{pc~cm^{-3}}$.
    \item Events with a detected signal-to-noise ratio (bonsai\_snr) $>$ 10.
    \item Events with excluded\_flag = 0 in CHIME/FRB Catalog 1.
    \item Event with sub-burst number $=$ 0 in CHIME/FRB Catalog 1.
\end{enumerate}

The selection items 3-5 are only conducted for non-repeaters. After applying the selection criteria, we have 377 non-repeaters and 59 repeaters selected. Assuming that the distribution of FRB sources is uniform and isotropic in the cosmos, the observed number should increase with the distance. However, due to the insensitivity to faint, high-DM FRBs, as well as additional unknown systematics of telescopes, the currently detected number of high-DM FRBs is incomplete, and this incompleteness increases with higher values of $\mathrm{DM_{ex}}$ \citep{2019NatAs...3..928R}.

\begin{figure}[htb!]
    \centering
    \includegraphics[width=0.8\textwidth]{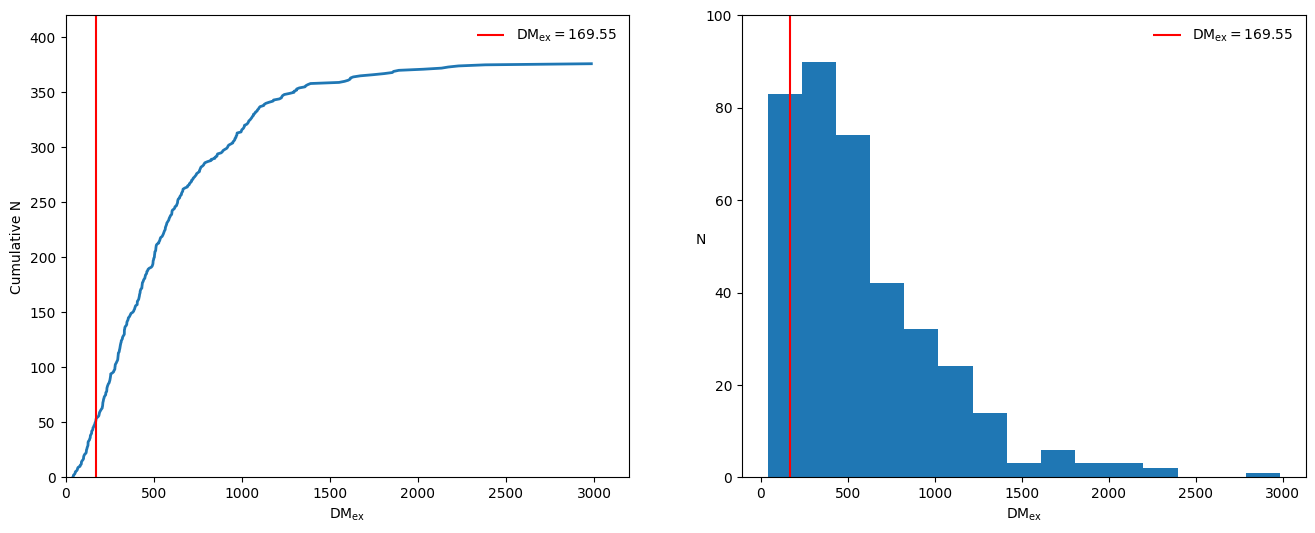}
    \caption{Cumulative distribution (left) and number distribution (right) of $\mathrm{DM_{ex}}$ for CHIME non-repeaters.}
    \label{fig:03}
\end{figure}

\begin{figure}[htb!]
    \centering
    \includegraphics[width=0.8\textwidth]{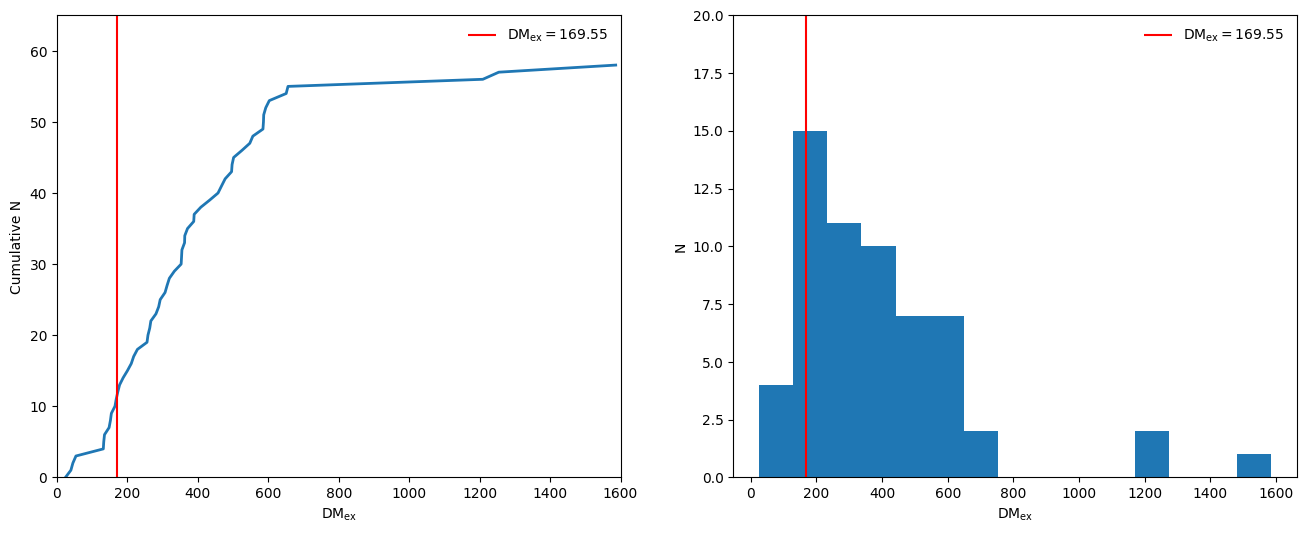}
    \caption{Cumulative distribution (left) and number distribution (right) of $\mathrm{DM_{ex}}$ for CHIME repeaters.}
    \label{fig:04}
\end{figure}

We plotted a histogram and cumulative number distribution for the selected samples in Figures \ref{fig:03} and \ref{fig:04}. As we mentioned before, both non-repeaters and repeaters show a lack of numbers in the high-DM region. So we do not estimate the event rate over a large space and instead focus on the local volumetric rate estimation for both non-repeaters and repeaters.

\section{Results}
\label{sec:results}

\subsection{Volumetric Rate Comparison}
\label{rate_comparison}

\begin{figure}[htb!]
    \centering
    \includegraphics[width=0.8\textwidth]{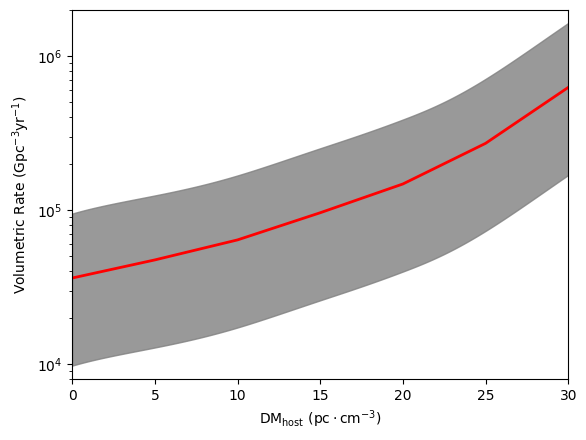}
    \caption{Volumetric rate (red line) estimated for different $\mathrm{DM_{host}}$. The rate is estimated using three non-repeaters with the lowest $\mathrm{DM_{ex}}$ for comparison with \citet{2019NatAs...3..928R}. The 95\% confidence region is shown as gray.}
    \label{fig:05}
\end{figure}

In this subsection, we display the result estimated by three non-repeaters with the lowest $\mathrm{DM_{ex}}$ for comparison with \citet{2019NatAs...3..928R}. By choosing these samples, we found that $\mathrm{DM_{ex}}<50~\mathrm{pc~cm^{-3}}$ for them, which is smaller than the samples from \citet{2019NatAs...3..928R} indicating a small $\mathrm{DM_{host}}$. Assuming no $\mathrm{DM_{host}}$ contribution, we get $d_\mathrm{limit} = 487~\mathrm{Mpc}$, and $R\approx 3.6^{+5.9}_{-2.6}\times10^4~\mathrm{Gpc^{-3}~yr^{-1}}$. The estimated event rates (red line) as well as the 95\% confidence limit (gray region) are shown in Figure \ref{fig:05}. The volumetric rate increases rapidly as the $\mathrm{DM_{host}}$ increases, which is because of the uncertainty from $\mathrm{DM_{IGM}}$ and $\mathrm{DM_{host}}$. For $\mathrm{DM_{host}}=0$, we find that the result is consistent with \citet{2019NatAs...3..928R}, indicating that the distribution of $\mathrm{DM_{IGM}}$ affects the rate estimation. Thus, more localized FRBs are needed to study the $\mathrm{DM_{IGM}}$ distribution. For local rate estimation, we choose samples that estimated redshift $z<0.2$ (corresponding to $\mathrm{DM_{ex}}<169.55~\mathrm{pc~cm^{-3}}$ with no $\mathrm{DM_{host}}$ assumption) considering the observational incompleteness and present the rate estimation for non-repeaters and repeaters in Sections \ref{rate_n} and \ref{rate_r}.

\subsection{Rate Estimation for Non-Repeaters}
\label{rate_n}
\begin{figure}[htb!]
    \centering
    \includegraphics[width=0.8\textwidth]{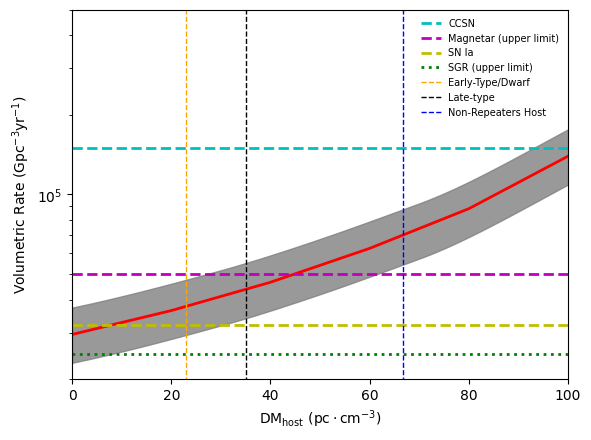}
    \hspace{0.1in}
    \caption{Event rate density estimation for different $\mathrm{DM_{host}}$ contribution assumptions of non-repeaters. The red solid line with a dark gray area represents the event rate at a $95\%$ confidence level. Host contributions are indicated by vertical lines, while rates of potential progenitors are shown as horizontal lines: SN Ia \citep{2010ApJ...713.1026D}, CCSN \citep{2008A&A...479...49B}, upper limit rate pf magnetars \citep{2008MNRAS.391.2009K}, and upper limit rate for SGR giant flares \citep{2007ApJ...659..339O, 2020MNRAS.494..665L}.}
    \label{fig:06}
\end{figure}

There are 51 non-repeaters that satisfied $\mathrm{DM_{ex}}<169.55~\mathrm{pc~cm^{-3}}$. We take effective observation time ($214.8$ days) as $t_\mathrm{obs}$ \citep{2021ApJS..257...59C} since all the non-repeaters are from CHIME/FRB Catalog 1. The rate for non-repeaters is $R\approx 6.9^{+1.82}_{-1.51}\times10^4~\mathrm{Gpc^{-3}~yr^{-1}}$, assuming the host galaxy contribution is the medium of localized non-repeaters $\mathrm{DM_{host}}=66.63~\mathrm{pc~{cm^{-3}}}$. The non-repeater rate is consistent with \citet{2023ApJ...944..105S}, who inferred a rate of $R=7.3^{+8.8}_{-3.8}(\mathrm{stat.})^{+2.0}_{-1.8}(\mathrm{sys.})\times10^{4}~\mathrm{{Gpc}^{-3}~{yr}^{-1}}$ above a pivot energy of $10^{39}$ erg and below a scattering timescale of $10$ ms at $600$ MHz. Figure \ref{fig:06} shows the estimated rate for non-repeaters under different $\mathrm{DM_{host}}$ assumptions. The red solid line and dark gray area represent the event rate at a $95\%$ confidence level. The vertical purple dashed line indicates the median of the host galaxy DM contribution estimated from localized non-repeaters samples in Section \ref{host}, and the vertical yellow and black dashed lines are the characteristic $\mathrm{DM_{host}}$ values for orientation-averaged early-type and dwarf galaxies, and for face-on late-type galaxies \citep{2019NatAs...3..928R}. We also compare the estimated rate with predictions for potential non-repeater progenitors. The estimated rate for SN Ia at $z=0.2$ is about $3.2\times10^4~\mathrm{Gpc^{-3}~yr^{-1}}$ \citep{2010ApJ...713.1026D}, the rate for core-collapse supernova (CCSN) is about $1.5\times 10^5~\mathrm{Gpc^{-3}~yr^{-1}}$ \citep{2008A&A...479...49B}, upper limit rate for magnetars is about $5\times 10^4~\mathrm{Gpc^{-3}~yr^{-1}}$ \citep{2008MNRAS.391.2009K}, and the upper limit rate for soft gamma-ray repeater (SGR) giant flares is about $2.5 \times 10^4~\mathrm{Gpc^{-3}~yr^{-1}}$ \citep{2007ApJ...659..339O, 2020MNRAS.494..665L}. The rate is compatible with the SN Ia rate and reaches the upper limit of magnetars and SGRs at the low $\mathrm{DM_{host}}$ scenario, but is larger than that rate when $\mathrm{DM_{host}}$ increases.

\subsection{Rate Estimation for Repeaters}
\label{rate_r}
\begin{figure}[htb!]
    \centering
    \hspace{0.1in}
    \includegraphics[width=0.8\textwidth]{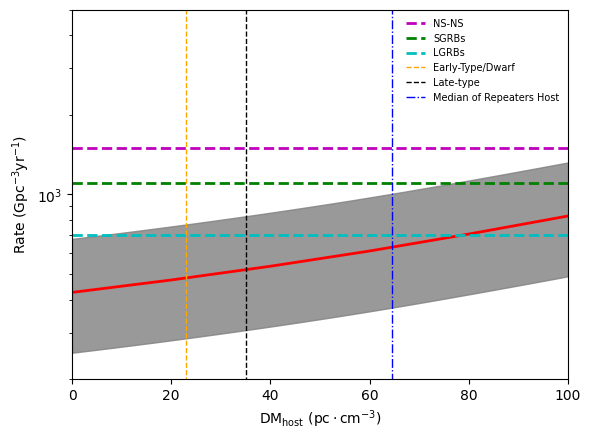}
    \caption{Event rate density estimation for different $\mathrm{DM_{host}}$ contribution assumptions of repeaters. The red solid line with a dark gray area represents the event rate at a $95\%$ confidence level. The rate of LGRBs \citep{2007MNRAS.382L..21C, 2020MNRAS.494..665L}, SGRBs \citep{2012MNRAS.425.2668C, 2020MNRAS.494..665L} and NS-NS mergers estimated from GW detection \citep{PhysRevLett.119.161101} are shown as horizontal lines. Host contributions are indicated by vertical lines, too.}
    \label{fig:07}
\end{figure}

Due to the rarity of repeater sources and their unclear repeating nature, we consider bursts from individual sources as one repeater without discussing the burst rate. We selected 13 sources in total (here, we exclude FRB 20200120E). We take $\tau = 4~\mathrm{yr}$ operating period of CHIME before 2024 as the observation duration to roughly estimate the event rate since the observational time of repeaters is difficult to calculate. Figure \ref{fig:07} shows the estimated rate density for repeaters. The red solid line with the dark gray area represents the event rate at a $95\%$ confidence level, assuming a Poisson event. The median of the $\mathrm{DM_{host}}$ contribution estimated using localized repeater sources is shown as the vertical orange dashed line. Assuming $\mathrm{DM_{host}}=64.46~\mathrm{pc~{cm^{-3}}}$, which is the median of the host contribution estimated from localized repeaters samples in Section \ref{host}, the rate for repeaters is $R\approx 630^{+375}_{-257}~\mathrm{Gpc^{-3}~yr^{-1}}$. FRB events may have a connection with both long and short GRB events in the collapse of supermassive neutron stars (SMNS) and may produce FRBs hundreds to thousands of seconds after GRBs \citep{2014ApJ...780L..21Z, 2023RvMP...95c5005Z}, \citet{2017ApJ...841...14M} also indicated the potential connection between GRBs and dwarf galaxies from FRB 20121102A. So we compare the repeater rate with the rate of LGRBs ($R\approx 7\times 10^2~\mathrm{Gpc^{-3}~yr^{-1}}$) and SGRBs ($R\approx 1.1\times 10^3~\mathrm{Gpc^{-3}~yr^{-1}}$) \citep{2007MNRAS.382L..21C, 2012MNRAS.425.2668C, 2020MNRAS.494..665L}. Furthermore, one suggests that repeaters are produced by binary systems (i.e. BWD mergers, binary neutron star (BNS) mergers, NS-BH mergers, etc.). Here, we only present the rate of BNS mergers from gravitational-wave observation ($R\approx 1.5\times 10^3~\mathrm{Gpc^{-3}~yr^{-1}}$) \citep{PhysRevLett.119.161101} in Figure \ref{fig:07}. For repeaters, the rate is consistent with LGRBs and SGRBs considering the $\mathrm{DM_{host}}$ contribution.

\section{Conclusions and Discussion}
\label{sec:discussion}

In this paper, we firstly compare the effects of the event rate for different $\mathrm{DM_{IGM}}$ distributions, and find that using different IGM distributions will directly affect the distance estimation of FRBs, as well as the event rate estimation. Secondly, we organize and compare the information about currently localized FRB host galaxies or candidates. Our results show that the inferred $\mathrm{DM_{host}}$ contributions of both localized repeaters and non-repeaters exhibit a log-normal distribution similar to \citet{2020ApJ...900..170Z} using Bayesian analysis. Based on the analysis of limited samples, the $\mathrm{DM_{host}}$ contributions of the two populations appear similar, suggesting that their origin environments may be somewhat consistent. However, the similarity is complex and could be due to factors such as limited samples, host galaxy alignment, and source offset within galaxies. Finally, we focus on the local volumetric rate for non-repeaters and repeaters under different $\mathrm{DM_{host}}$ assumptions. We analyze samples of repeaters and non-repeaters observed by the CHIME telescope, and estimate the impact of different $\mathrm{DM_{host}}$ contributions on volumetric event rates. \citet{2019NatAs...3..928R} estimated the lower limit of event rates using non-repeaters, while \citet{2020ApJ...893...44Z} introduced a new model to estimate the event rate of repeaters from BNS mergers, indicating few events from this scenario. We excluded FRBs with extremely low $\mathrm{DM_{ex}}$ to avoid contamination and excluded those with large DM due to observational incompleteness. It turns out that the contribution of $\mathrm{DM_{host}}$ may significantly influence local volumetric rates, suggesting complicated FRB origins.

To calculate $\mathrm{DM_{ex}}$, we subtract contributions from $\mathrm{DM_{MW}}$ and $\mathrm{DM_{halo}}$ using the YMW16 model and YT20 model. Using the PyGEDM package to calculate DM contributions with these models in a Python environment is very convenient. We confirm that our results are reasonable, with the final results obtained using the NE2001 model for Galactic DM contributions consistent with existing results. However, it is confusing that the $\mathrm{DM_{ex}}$ and $\mathrm{DM_{host}}$ of several FRBs are too small (even less than 0). For instance, FRB 20180916B ($\mathrm{DM_{obs}=349.00~pc~{cm}^{-3}}$) was localized to a spiral galaxy with $z\approx 0.0337$. The contributions estimated by the two electron density models are $\mathrm{DM_{YMW16}=324.82~pc~{cm}^{-3}}$ and $\mathrm{DM_{NE2001}=198.91~pc~{cm}^{-3}}$, respectively. Clearly, the $\mathrm{DM_{ex}}$ calculated using YMW16 is too small, inconsistent with its cosmological origin. For FRB 20180814A ($\mathrm{DM_{obs}=189.40~pc~{cm}^{-3}}$), which $\mathrm{DM_{ex}=45.32~pc~{cm}^{-3}}$ was calculated from the YMW16 and YT20 models. It has been localized to a galaxy with $z=0.0684$ and the value of $\mathrm{DM_{host}}$ becomes negative estimated from the $\mathrm{DM_{IGM}}-z$ relation. One possible reason is that the YMW16 model overestimates the DM contributions from the Milky Way along the lines of sight, leading to inconsistencies with observations, and another reason is the uncertainties in the $\mathrm{DM_{IGM}}-z$ relation.

The morphologies of the host galaxy are currently incomplete and unclear. About half of host galaxies are recognized as spiral galaxies, with few identified as dwarf galaxies. The analysis of the estimated host galaxy contributions for currently localized FRBs shows no significant difference between repeaters and non-repeaters. However, the $\mathrm{DM_{host}}$ depends on the direction from which we observe FRBs, more localized FRBs are needed to average the impact from orientation. Besides, there is evidence that several FRBs originate from complicated environments \citep{2022Natur.606..873N, 2022Natur.609..685X, 2023Sci...380..599A}, and the total stellar mass and SFR of host galaxies show variety, too, which indicates that FRBs have multiple origins. Discussing the host galaxy properties between non-repeaters and repeaters might not be reasonable due to the unclear classification of FRBs. We suggest reclassifying the FRBs by different aspects or subclassifying the non-repeaters and repeaters \citep{2022JCAP...07..010G}.

By using the analytical form of $\mathrm{DM_{IGM}}$, we found a volumetric rate $R\approx 3.6^{+5.9}_{-2.6}\times10^4~\mathrm{Gpc^{-3}~yr^{-1}}$ under the assumption of $\mathrm{DM_{host}}=0$. This rate is higher than that of \citet{2019NatAs...3..928R} despite the FRBs we chose having smaller $\mathrm{DM_{ex}}$. However, there is no expectation of a large $\mathrm{DM_{host}}$ for these sources, but the rate will be affected significantly due to the change of $\mathrm{DM_{host}}$. For example, the rate assuming $\mathrm{DM_{host}}=30~\mathrm{pc~cm^{-3}}$ is higher than the rate with no $\mathrm{DM_{host}}$ by a factor of 10. In Section \ref{rate_n} and \ref{rate_r}, we estimate the volumetric rate for non-repeaters and repeaters using FRBs with $z<0.2$ and compare them with potential progenitors. We cannot exclude any origin models from our analysis. Noticed that the observation time in Equation (\ref{eq:08}) can only be roughly estimated, especially for repeaters, thus the actual volumetric event rate might be higher than our estimated result. In addition, some researchers predict that non-repeaters might eventually repeat, and the ratio of repeaters to non-repeaters may change. 

We do not discuss the luminosity/energy function, such as \citet{2018MNRAS.481.2320L, 2020MNRAS.494..665L}, \citet{2022MNRAS.511.1961H}, \citet{2023ApJ...944..105S} did, to determine the event rate density of an FRB. One reason is the observational incompleteness in the high-DM region, so we only focus on the local volumetric rate density. Besides, the unknown $\mathrm{DM_{host}}$ contribution and the redshift will result in uncertain luminosity or energy. Therefore, the redshift evolution of different populations is also unclear. Observationally, the non-repeaters typically have higher luminosities or energies than repeaters but the number of non-repeaters is larger than that of repeaters. This inconsistency with the results of luminosity/energy function suggests that the low-luminosity FRBs ($L<10^{43}~\mathrm{erg~s}^{-1}$) could be repeater candidates \citep{2020MNRAS.494..665L}.

The origin of FRBs remains inconclusive. Many models are proposed for repeating and non-repeating FRBs. Moreover, some non-repeaters arising from progenitors can generate repeaters essentially. \citet{2020ApJ...893...44Z} suggest that the repeater rate from magnetars formed in BNS mergers ranges from hundreds to thousands, depending on their active lifetimes. GRBs may originate from two types of progenitors: the massive star collapse and NS-NS or NS-BH coalescence. Only a small fraction of FRBs could be associated with GRBs. LGRBs exhibit weak radio signals due to the more isotropic supernova, while SGRBs may be transparent \citep{2014ApJ...780L..21Z}. We need to test the redshift distribution and evolution for different populations. Although several studies on the redshift evolution of FRB rates suggest that FRBs may not evolve with the SFR, it cannot be excluded that some FRBs are associated with such events. Population synthesis on a small sample of Parkes and ASKAP bursts indicates evolution with SFR \citep{2022MNRAS.510L..18J}, while recent research shows different evolution patterns between FRBs and SFR \citep{2022MNRAS.511.1961H, 2022ApJ...924L..14Z, 2023Univ....9..251Z}. Newborn neutron stars are mainly produced by the CCSN, and the formation rate may trace the SFR. However, the old stellar populations might cause significant time delay with respect to star formation and track the total stellar mass. A hybrid model of young and old stellar populations is also reasonable.

In the future, an increasing number of FRBs will be detected across a wide range of DM values. This will allow for a more detailed examination of local event rates, better constraints on origins, and the study of the redshift distribution and evolution of FRBs. By comparing the properties of FRB hosts with different progenitors and other astrophysical transients, we can constrain origin models more efficiently. For example, we can compare source positions and host galaxy metallicities to those of ultraluminous X-ray sources, which tend to reside near young stellar clusters in low metallicity hosts \citep{2021ApJ...917...13S}, and GRBs, which have attractive properties as well as FRBs. Furthermore, neutron stars, particularly magnetars, should be further explored as prominent models for FRBs, including discussions on BNS mergers detectable through gravitational radiation \citep{PhysRevLett.119.161101}. An analysis of the most recent Gravitational Wave Transient Catalog 3 has inferred a merger rate ranging from 10 to 1700 $\mathrm{Gpc^{-3}~yr^{-1}}$ for BNS and from 7.8 to 140~$\mathrm{Gpc^{-3}~yr^{-1}}$ for NSBH populations. Based on our results, the rate of repeaters is consistent with that could possibly be associated with BNS. As noted by \citet{2020MNRAS.494..665L}, if BNS and/or NSBH sources were only associated with FRBs from the high end of the luminosity function ($\ge 10^{43}~\mathrm{erg~s^{-1}}$), such rates could be comparable. The Advanced LIGO/Virgo observations can discover more potential connections between FRBs and gravitational-wave events in collaboration with CHIME/FRB observations \citep{2022ApJ...937...89W, 2023ApJ...955..155A}.



\section*{Acknowledgements}

\noindent This work was supported by the National Science Foundation of Xinjiang Uygur Autonomous Region (2022D01D85), the CAS Project for Young Scientists in Basic Research (YSBR-063), the Major Science and Technology Program of Xinjiang Uygur Autonomous Region (Nos. 2022A03013-1, 2022A03013-2), the Tianshan talents program (2023TSYCTD0013),  and Ju-Mei Yao was supported by the Tianchi Talent project. The work was also supported by the National SKA Program of China/2022SKA0130100, the Office of the Leading Group for Cyberspace Affairs, CAS (No. CAS-WX2023PY-0102), the CAS Youth Interdisciplinary Team, the Guizhou Provincial Education Department (grant No. KY(2023)059), the National Science Foundation of China (Nos. 12288102, 12273009, 11988101, 12203069, 12041302, 12203045) and the Chinese Academy of Sciences (CAS) “Light of West China” Program (No. 2022-XBQNXZ-015).

\appendix

\section{Table of Localized FRBs and Their Host Galaxy Properties}
\label{appendix:A}

Here, we present comprehensive information about the localized FRBs used in our research in Table \ref{chartable:01}, as well as stellar mass, SFR and inferred galaxy morphology of their host galaxies. We set galaxy morphology as “unclear” if there is no mention of it in the reference article.

\begin{longrotatetable}
\movetabledown=0.5in
\begin{deluxetable*}{lllllllCccl}
\tablecaption{Properties of Localized Non-repeaters and Host Galaxies\label{chartable:01}}
\tablewidth{700pt}
\tabletypesize{\scriptsize}
\tablehead{
\colhead{Source} & \colhead{Telescope} & 
\colhead{Repeater} & \colhead{GL} & 
\colhead{GB} & \colhead{Morphology} & 
\colhead{Stellar Mass} & \colhead{SFR} & 
\colhead{Redshift} & \colhead{$\mathrm{DM_{obs}}$} & \colhead{References} \\ 
\colhead{} & \colhead{} & \colhead{} & \colhead{(deg)} & 
\colhead{(deg)} & \colhead{} & \colhead{($M_\odot$)} &
\colhead{($M_\odot~\mathrm{yr}^{-1}$)} & \colhead{} & \colhead{($\mathrm{pc~cm^{-3}}$)} & \colhead{}
}
\startdata
FRB 20171020A & ASKAP & No & 29.3 & -51.3 & Spiral & $9\times10^8$ & $0.13$ & 0.0087 & 114.10 & \citet{2018ApJ...867L..10M} \\
FRB 20180924B & ASKAP & No & 0.74 & -49.41 & Spiral(early type) & $2.2\times10^{10}$ & $<2.0$  & 0.3214 & 361.42 & \citet{2019Sci...365..565B}\\
FRB 20181112A & ASKAP & No & 342.6 & -47.7 & Spiral(star-forming) & $2.6\times10^9$ & $0.6$ & 0.4755 & 589.27 & \citet{2019Sci...366..231P} \\
FRB 20190102C & ASKAP & No & 312.65 & -33.49 & Spiral(star-forming) & $10^{9.5}$ & $1.5$ & 0.2913 & 363.60 & \citet{2020ApJ...895L..37B} \\
FRB 20190608B & ASKAP & No & 53.21 & -48.53 & Spiral(star-forming) & $10^{10.4}$ & $1.2$ & 0.1178 & 338.70 & \citet{2020ApJ...895L..37B} \\
FRB 20190611B & ASKAP & No & 312.94 & -33.28 & Unclear(no spiral arms) & $0.8\times10^9$ & $0.27$ & 0.3778 & 321.40 & \citet{2020ApJ...903..152H} \\
FRB 20190714A & ASKAP & No & 289.7 & 49.0 & Spiral structure & $14.9\times10^9$ & $0.65$ & 0.2365 & 504.00 & \citet{2020ApJ...903..152H}, \citet{2021ApJ...917...75M} \\
FRB 20191001A & ASKAP & No & 341.3 & -44.8 & Spiral structure & $46.4\times10^9$ & $8.06$ & 0.2340 & 506.92 & \citet{2020ApJ...903..152H}, \citet{2021ApJ...917...75M} \\
FRB 20191228A & ASKAP & No & 20.8 & -64.9 & Unclear & $0.54\times10^{10}$ & $0.50$ & 0.2430 & 297.5 & \citet{2022AJ....163...69B} \\
FRB 20200430A & ASKAP & No & 17.06 & 52.52 & Unclear & $1.3\times10^9$ & $0.2$ & 0.1600 & 380.1 & \citet{2020ApJ...903..152H} \\
FRB 20200906A & ASKAP & No & 202.4 & -49.77 & Unclear & $1.33\times10^{10}$ & $0.48$ & 0.3688 & 577.8 & \citet{2022AJ....163...69B} \\
FRB 20210117A & ASKAP & No & 45.79 & -57.59 & Dwarf & $10^{8.6}$ & $0.014$ & 0.2140 & 729.10 & \citet{2023ApJ...948...67B} \\
FRB 20210320C & ASKAP & No & 318.90 & 45.30 & Unclear & $10^{10.37}$ & $3.51$ & 0.2797 & 384.80 & \citet{2023ApJ...954...80G}, \citet{2022MNRAS.516.4862J} \\
FRB 20210807D & ASKAP & No & 39.81 & -14.89 & Quiescent & $10^{10.97}$ & $0.63$ & 0.1293 & 651.50 & \citet{2023ApJ...954...80G} \\
FRB 20211127I & ASKAP & No & 311.99 & 43.56 & Spiral & $3\times10^9$ & $0.45$ & 0.0469 & 234.83 & \citet{2023ApJ...949...25G} \\
FRB 20211203C & ASKAP & No & 314.43 & 30.47 & Unclear & $10^{9.76}$ & $15.91$ & 0.3437 & 636.20 & \citet{2023ApJ...954...80G}\\
FRB 20211212A & ASKAP & No & 243.95 & 47.76 & Spiral & $10^{10.28}$ & $0.73$ & 0.0707 & 206.00 & \citet{2023ApJ...954...80G} \\
FRB 20220105A & ASKAP & No & 18.84 & 74.68 & Unclear & $10^{10.01}$ & $0.42$ & 0.2784 & 583.00 & \citet{2023ApJ...954...80G}\\
FRB 20220610A & ASKAP & No & 8.87 & -70.13 & Spiral & $10^{9.98}$ & $0.42$ & 1.0170 & 1458.10 & \citet{2023Sci...382..294R} \\
FRB 20230718A & ASKAP & No & 259.66 & -1.03 & Spiral(star-forming) & $8.28\times10^8$ & $1.69$ & 0.0359 & 477.00 & \citet{2023ApJ...949...25G} \\
FRB 20181220A & CHIME & No & 105.24 & -10.73 & Spiral & $10^{9.86}$ & $3.0$ & 0.0275 & 209.40 & \citet{2024ApJ...971L..51B} \\
FRB 20181223C & CHIME & No & 207.75 & 79.51 & Spiral(star-forming) & $10^{9.29}$ & $0.054$ & 0.0302 & 112.50 & \citet{2024ApJ...971L..51B} \\
FRB 20190418A & CHIME & No & 179.30 & -22.93 & Spiral & $10^{10.27}$ & $0.15$ & 0.0715 & 184.50 & \citet{2024ApJ...971L..51B} \\
FRB 20190425A & CHIME & No & 42.06 & 33.02 & Spiral(star-forming) & $10^{10.26}$ & $1.5$ & 0.0312 & 128.20 & \citet{2024ApJ...971L..51B} \\
FRB 20210603A & CHIME & No & 119.71 & -41.58 & Unclear & $8.5\times10^{10}$ & $0.24$ & 0.1770 & 500.15 & \citet{2024NatAs...8.1429C} \\
FRB 20190523A & DSA-110 & No & 117.03 & 44.0 & Unclear & $10^{11.07}$ & $<1.3$ & 0.6600 & 760.80 & \citet{2019Natur.572..352R} \\
FRB 20220207C & DSA-110 & No & 106.94 & 18.39 & Disk dominated & $10^{9.91}$ & $1.16$ & 0.0430 & 263.00 & \citet{2024ApJ...967...29L} \\
FRB 20220307B & DSA-110 & No & 116.24 & 10.47 & Unclear & $10^{10.04}$ & $2.71$ & 0.2481 & 499.328 & \citet{2024ApJ...967...29L} \\
FRB 20220310F & DSA-110 & No & 140.02 & 34.80 & Unclear & $10^{9.97}$ & $4.25$ &  0.4780 & 462.657 & \citet{2024ApJ...967...29L} \\
FRB 20220319D & DSA-110 & No & 129.18 & 9.11 & Spiral & $10^{10.16}$ & $0.21$ & 0.0112 & 110.95 & \citet{2024ApJ...967...29L}, \citet{2023arXiv230101000R} \\
FRB 20220418A & DSA-110 & No & 110.75 & 44.47 & Unclear & $10^{10.31}$ & $29.41$ & 0.6220 & 624.124 & \citet{2024ApJ...967...29L} \\
FRB 20220506D & DSA-110 & No & 108.35 & 16.51 & Unclear & $10^{10.29}$ & $5.08$ & 0.3004 & 396.651 & \citet{2024ApJ...967...29L} \\
FRB 20220509G & DSA-110 & No & 100.94 & 25.48 & Spiral(early type) & $10^{10.79}$ & $0.23$ & 0.0894 & 270.26 & \citet{2024ApJ...967...29L}, \citet{2023ApJ...950..175S} \\
FRB 20220825A & DSA-110 & No & 106.99 & 17.79 & Unclear & $10^{9.95}$ & $1.62$ & 0.2414 & 649.893 & \citet{2024ApJ...967...29L} \\
FRB 20220914A & DSA-110 & No & 104.31 & 26.13 & Star-forming(late type) & $10^{9.48}$ & $0.72$ & 0.1139 & 630.703 & \citet{2024ApJ...967...29L} \\
FRB 20220920A & DSA-110 & No & 104.92 & 38.89 & Unclear & $10^{9.81}$ & $4.10$ & 0.1582 & 314.98 & \citet{2024ApJ...967...29L} \\
FRB 20221012A & DSA-110 & No & 101.14 & 26.14 & Unclear & $10^{10.99}$ & $0.15$ & 0.2847 & 440.36 & \citet{2024ApJ...967...29L} \\
FRB 20210405I & MeerKAT & No & 338.19 & -4.59 & Spiral(early type) & $10^{11.25}$ & $\ge 0.3$ & 0.0660 & 566.43 & \citet{2024MNRAS.527.3659D} \\
FRB 20210410D & MeerKAT & No & 312.32 & -34.13 & Dwarf & $10^{9.46}$ & $0.03$ & 0.1415 & 578.78 & \citet{2023MNRAS.524.2064C} \\
FRB 20150418A & Parkes & No & 232.67 & -3.23 & Elliptical & $10^{11}$ & $<0.2$ & 0.4920 &  776.2 & \citet{2016Natur.530..453K} \\
FRB 20190614D & VLA & No & 136.3 & 16.5 & Unclear & $10^{9.6}$ & $>0.0$ & 0.6000 & 959.20 & \citet{2020ApJ...899..161L} \\
\hline
FRB 20121102A & Arecibo & Yes & 174.89 & -0.23 & Dwarf & $10^8$ & $0.23$ & 0.1927 & 557.40 & \citet{2017ApJ...834L...7T} \\
FRB 20190711A & ASKAP & Yes & 310.91 & -33.9 & Unclear & $0.81\times10^9$ & $0.42$ & 0.5220 & 593.10 & \citet{2020ApJ...903..152H} \\
FRB 20180814A & CHIME & Yes & 136.46 & 16.58 & Spiral(early type) & $10^{10.78}$ & $\le 0.32$ & 0.0684 & 189.40 & \citet{2023ApJ...950..134M} \\
FRB 20180916B & CHIME & Yes & 129.71 & 3.73 & Spiral(star-forming) & $2.74\times10^9$ & $>0.016$ & 0.0337 & 349.00 & \citet{2020Natur.577..190M}, \citet{2022ApJ...925L..20K} \\
FRB 20181030A & CHIME & Yes & 133.4 & 40.9 & Spiral(star-forming) & $5.8\times10^9$ & $0.36$ & 0.0039 & 103.50 & \citet{2021ApJ...919L..24B} \\
FRB 20190110C & CHIME & Yes & 65.52 & 43.85 & Star-forming & $2.5\times10^{10}$ & $0.54$ & 0.1224 & 221.96 & \citet{2024ApJ...961...99I} \\
FRB 20190303A & CHIME & Yes & 97.5 & 65.7 & Spiral(star-forming) & $10^{10.63}$ & $9.77$ & 0.0640 & 222.40 & \citet{2023ApJ...950..134M} \\
FRB 20191106C & CHIME & Yes & 105.70 & 73.20 & Elliptical & $45\times10^{10}$ & $4.75$ & 0.1078 & 333.40 & \citet{2024ApJ...961...99I} \\
FRB 20200120E & CHIME & Yes & 142.19 & 41.22 & Spiral(globular cluster) & $7.2\times10^{10}$ & $0.6$ & 0.0008 & 87.82 & \citet{2021ApJ...910L..18B}, \citet{2022Natur.602..585K} \\
FRB 20200223B & CHIME & Yes & 118.10 & -33.90 & Spiral & $5.6\times10^{10}$ & $0.59$ & 0.0602 & 202.27 & \citet{2024ApJ...961...99I} \\
FRB 20201124A & CHIME & Yes & 177.60 & -8.50 & Spiral & $2.5\times10^{10}$ & $3.4$ & 0.0980 & 410.83 & \citet{2022Natur.609..685X}, \citet{2022MNRAS.513..982R} \\
FRB 20220912A & CHIME & Yes & 347.27 & 48.70 & Disk dominated(late type) & $10^{10}$ & $\ge 0.1$ & 0.0771 & 219.46 & \citet{2023ApJ...949L...3R} \\
FRB 20190520B & FAST & Yes & 359.67 & 29.91 & Dwarf & $6\times10^8$ & $0.41$ & 0.2410 & 1204.7 & \citet{2022Natur.606..873N} \\
FRB 20180301A & Parkes & Yes & 204.412 & -6.481 & Unclear & $0.23\times10^{10}$ & $1.93$ & 0.3304 & 522.00 & \citet{2022AJ....163...69B} \\
\enddata
\end{deluxetable*}
\end{longrotatetable}


\bibliography{sample631}{}

\begin{thebibliography}{}
\expandafter\ifx\csname natexlab\endcsname\relax\def\natexlab#1{#1}\fi
\providecommand{\url}[1]{\href{#1}{#1}}
\providecommand{\dodoi}[1]{doi:~\href{http://doi.org/#1}{\nolinkurl{#1}}}
\providecommand{\doeprint}[1]{\href{http://ascl.net/#1}{\nolinkurl{http://ascl.net/#1}}}
\providecommand{\doarXiv}[1]{\href{https://arxiv.org/abs/#1}{\nolinkurl{https://arxiv.org/abs/#1}}}

\bibitem[{Abbott {et~al.}(2017)Abbott, Abbott, Abbott, Acernese, Ackley, Adams, Adams, Addesso, Adhikari, Adya, Affeldt, Afrough, Agarwal, Agathos, Agatsuma, Aggarwal, Aguiar, Aiello, Ain, Ajith, Allen, Allen, Allocca, Altin, Amato, Ananyeva, Anderson, Anderson, Angelova, Antier, Appert, Arai, Araya, Areeda, Arnaud, Arun, Ascenzi, Ashton, Ast, Aston, Astone, Atallah, Aufmuth, Aulbert, AultONeal, Austin, Avila-Alvarez, Babak, Bacon, Bader, Bae, Bailes, Baker, Baldaccini, Ballardin, Ballmer, Banagiri, Barayoga, Barclay, Barish, Barker, Barkett, Barone, Barr, Barsotti, Barsuglia, Barta, Barthelmy, Bartlett, Bartos, Bassiri, Basti, Batch, Bawaj, Bayley, Bazzan, B\'ecsy, Beer, Bejger, Belahcene, Bell, Berger, Bergmann, Bernuzzi, Bero, Berry, Bersanetti, Bertolini, Betzwieser, Bhagwat, Bhandare, Bilenko, Billingsley, Billman, Birch, Birney, Birnholtz, Biscans, Biscoveanu, Bisht, Bitossi, Biwer, Bizouard, Blackburn, Blackman, Blair, Blair, Blair, Bloemen, Bock, Bode, Boer, Bogaert, Bohe, Bondu, Bonilla, Bonnand,
  Boom, Bork, Boschi, Bose, Bossie, Bouffanais, Bozzi, Bradaschia, Brady, Branchesi, Brau, Briant, Brillet, Brinkmann, Brisson, Brockill, Broida, Brooks, Brown, Brown, Brunett, Buchanan, Buikema, Bulik, Bulten, Buonanno, Buskulic, Buy, Byer, Cabero, Cadonati, Cagnoli, Cahillane, Calder\'on~Bustillo, Callister, Calloni, Camp, Canepa, Canizares, Cannon, Cao, Cao, Capano, Capocasa, Carbognani, Caride, Carney, Carullo, Casanueva~Diaz, Casentini, Caudill, Cavagli\`a, Cavalier, Cavalieri, Cella, Cepeda, Cerd\'a-Dur\'an, Cerretani, Cesarini, Chamberlin, Chan, Chao, Charlton, Chase, Chassande-Mottin, Chatterjee, Chatziioannou, Cheeseboro, Chen, Chen, Chen, Cheng, Chia, Chincarini, Chiummo, Chmiel, Cho, Cho, Chow, Christensen, Chu, Chua, Chua, Chung, Chung, Ciani, Ciolfi, Cirelli, Cirone, Clara, Clark, Clearwater, Cleva, Cocchieri, Coccia, Cohadon, Cohen, Colla, Collette, Cominsky, Constancio, Conti, Cooper, Corban, Corbitt, Cordero-Carri\'on, Corley, Cornish, Corsi, Cortese, Costa, Coughlin, Coughlin, Coulon,
  Countryman, Couvares, Covas, Cowan, Coward, Cowart, Coyne, Coyne, Creighton, Creighton, Cripe, Crowder, Cullen, Cumming, Cunningham, Cuoco, Dal~Canton, D\'alya, Danilishin, D'Antonio, Danzmann, Dasgupta, Da~Silva~Costa, Dattilo, Dave, Davier, Davis, Daw, Day, De, DeBra, Degallaix, De~Laurentis, Del\'eglise, Del~Pozzo, Demos, Denker, Dent, De~Pietri, Dergachev, De~Rosa, DeRosa, De~Rossi, DeSalvo, de~Varona, Devenson, Dhurandhar, D\'{\i}az, Dietrich, Di~Fiore, Di~Giovanni, Di~Girolamo, Di~Lieto, Di~Pace, Di~Palma, Di~Renzo, Doctor, Dolique, Donovan, Dooley, Doravari, Dorrington, Douglas, Dovale~\'Alvarez, Downes, Drago, Dreissigacker, Driggers, Du, Ducrot, Dudi, Dupej, Dwyer, Edo, Edwards, Effler, Eggenstein, Ehrens, Eichholz, Eikenberry, Eisenstein, Essick, Estevez, Etienne, Etzel, Evans, Evans, Factourovich, Fafone, Fair, Fairhurst, Fan, Farinon, Farr, Farr, Fauchon-Jones, Favata, Fays, Fee, Fehrmann, Feicht, Fejer, Fernandez-Galiana, Ferrante, Ferreira, Ferrini, Fidecaro, Finstad, Fiori, Fiorucci,
  Fishbach, Fisher, Fitz-Axen, Flaminio, Fletcher, Fong, Font, Forsyth, Forsyth, Fournier, Frasca, Frasconi, Frei, Freise, Frey, Frey, Fries, Fritschel, Frolov, Fulda, Fyffe, Gabbard, Gadre, Gaebel, Gair, Gammaitoni, Ganija, Gaonkar, Garcia-Quiros, Garufi, Gateley, Gaudio, Gaur, Gayathri, Gehrels, Gemme, Genin, Gennai, George, George, Gergely, Germain, Ghonge, Ghosh, Ghosh, Ghosh, Giaime, Giardina, Giazotto, Gill, Glover, Goetz, Goetz, Gomes, Goncharov, Gonz\'alez, Gonzalez~Castro, Gopakumar, Gorodetsky, Gossan, Gosselin, Gouaty, Grado, Graef, Granata, Grant, Gras, Gray, Greco, Green, Gretarsson, Groot, Grote, Grunewald, Gruning, Guidi, Guo, Gupta, Gupta, Gushwa, Gustafson, Gustafson, Halim, Hall, Hall, Hamilton, Hammond, Haney, Hanke, Hanks, Hanna, Hannam, Hannuksela, Hanson, Hardwick, Harms, Harry, Harry, Hart, Haster, Haughian, Healy, Heidmann, Heintze, Heitmann, Hello, Hemming, Hendry, Heng, Hennig, Heptonstall, Heurs, Hild, Hinderer, Ho, Hoak, Hofman, Holt, Holz, Hopkins, Horst, Hough, Houston, Howell,
  Hreibi, Hu, Huerta, Huet, Hughey, Husa, Huttner, Huynh-Dinh, Indik, Inta, Intini, Isa, Isac, Isi, Iyer, Izumi, Jacqmin, Jani, Jaranowski, Jawahar, Jim\'enez-Forteza, Johnson, Johnson-McDaniel, Jones, Jones, Jonker, Ju, Junker, Kalaghatgi, Kalogera, Kamai, Kandhasamy, Kang, Kanner, Kapadia, Karki, Karvinen, Kasprzack, Kastaun, Katolik, Katsavounidis, Katzman, Kaufer, Kawabe, K\'ef\'elian, Keitel, Kemball, Kennedy, Kent, Key, Khalili, Khan, Khan, Khan, Khazanov, Kijbunchoo, Kim, Kim, Kim, Kim, Kim, Kim, Kimbrell, King, King, Kinley-Hanlon, Kirchhoff, Kissel, Kleybolte, Klimenko, Knowles, Koch, Koehlenbeck, Koley, Kondrashov, Kontos, Korobko, Korth, Kowalska, Kozak, Kr\"amer, Kringel, Krishnan, Kr\'olak, Kuehn, Kumar, Kumar, Kumar, Kuo, Kutynia, Kwang, Lackey, Lai, Landry, Lang, Lange, Lantz, Lanza, Larson, Lartaux-Vollard, Lasky, Laxen, Lazzarini, Lazzaro, Leaci, Leavey, Lee, Lee, Lee, Lee, Lee, Lehmann, Lenon, Leon, Leonardi, Leroy, Letendre, Levin, Li, Linker, Littenberg, Liu, Liu, Lo, Lockerbie, London,
  Lord, Lorenzini, Loriette, Lormand, Losurdo, Lough, Lousto, Lovelace, L\"uck, Lumaca, Lundgren, Lynch, Ma, Macas, Macfoy, Machenschalk, MacInnis, Macleod, Maga\~na Hernandez, Maga\~na Sandoval, Maga\~na Zertuche, Magee, Majorana, Maksimovic, Man, Mandic, Mangano, Mansell, Manske, Mantovani, Marchesoni, Marion, M\'arka, M\'arka, Markakis, Markosyan, Markowitz, Maros, Marquina, Marsh, Martelli, Martellini, Martin, Martin, Martynov, Marx, Mason, Massera, Masserot, Massinger, Masso-Reid, Mastrogiovanni, Matas, Matichard, Matone, Mavalvala, Mazumder, McCarthy, McClelland, McCormick, McCuller, McGuire, McIntyre, McIver, McManus, McNeill, McRae, McWilliams, Meacher, Meadors, Mehmet, Meidam, Mejuto-Villa, Melatos, Mendell, Mercer, Merilh, Merzougui, Meshkov, Messenger, Messick, Metzdorff, Meyers, Miao, Michel, Middleton, Mikhailov, Milano, Miller, Miller, Miller, Millhouse, Milovich-Goff, Minazzoli, Minenkov, Ming, Mishra, Mitra, Mitrofanov, Mitselmakher, Mittleman, Moffa, Moggi, Mogushi, Mohan, Mohapatra, Molina,
  Montani, Moore, Moraru, Moreno, Morisaki, Morriss, Mours, Mow-Lowry, Mueller, Muir, Mukherjee, Mukherjee, Mukherjee, Mukund, Mullavey, Munch, Mu\~niz, Muratore, Murray, Nagar, Napier, Nardecchia, Naticchioni, Nayak, Neilson, Nelemans, Nelson, Nery, Neunzert, Nevin, Newport, Newton, Ng, Nguyen, Nguyen, Nichols, Nielsen, Nissanke, Nitz, Noack, Nocera, Nolting, North, Nuttall, Oberling, O'Dea, Ogin, Oh, Oh, Ohme, Okada, Oliver, Oppermann, Oram, O'Reilly, Ormiston, Ortega, O'Shaughnessy, Ossokine, Ottaway, Overmier, Owen, Pace, Page, Page, Pai, Pai, Palamos, Palashov, Palomba, Pal-Singh, Pan, Pan, Pang, Pang, Pankow, Pannarale, Pant, Paoletti, Paoli, Papa, Parida, Parker, Pascucci, Pasqualetti, Passaquieti, Passuello, Patil, Patricelli, Pearlstone, Pedraza, Pedurand, Pekowsky, Pele, Penn, Perez, Perreca, Perri, Pfeiffer, Phelps, Piccinni, Pichot, Piergiovanni, Pierro, Pillant, Pinard, Pinto, Pirello, Pitkin, Poe, Poggiani, Popolizio, Porter, Post, Powell, Prasad, Pratt, Pratten, Predoi, Prestegard, Prijatelj,
  Principe, Privitera, Prix, Prodi, Prokhorov, Puncken, Punturo, Puppo, P\"urrer, Qi, Quetschke, Quintero, Quitzow-James, Raab, Rabeling, Radkins, Raffai, Raja, Rajan, Rajbhandari, Rakhmanov, Ramirez, Ramos-Buades, Rapagnani, Raymond, Razzano, Read, Regimbau, Rei, Reid, Reitze, Ren, Reyes, Ricci, Ricker, Rieger, Riles, Rizzo, Robertson, Robie, Robinet, Rocchi, Rolland, Rollins, Roma, Romano, Romano, Romel, Romie, Rosi\ifmmode~\acute{n}\else \'{n}\fi{}ska, Ross, Rowan, R\"udiger, Ruggi, Rutins, Ryan, Sachdev, Sadecki, Sadeghian, Sakellariadou, Salconi, Saleem, Salemi, Samajdar, Sammut, Sampson, Sanchez, Sanchez, Sanchis-Gual, Sandberg, Sanders, Sassolas, Sathyaprakash, Saulson, Sauter, Savage, Sawadsky, Schale, Scheel, Scheuer, Schmidt, Schmidt, Schnabel, Schofield, Sch\"onbeck, Schreiber, Schuette, Schulte, Schutz, Schwalbe, Scott, Scott, Seidel, Sellers, Sengupta, Sentenac, Sequino, Sergeev, Shaddock, Shaffer, Shah, Shahriar, Shaner, Shao, Shapiro, Shawhan, Sheperd, Shoemaker, Shoemaker, Siellez, Siemens,
  Sieniawska, Sigg, Silva, Singer, Singh, Singhal, Sintes, Slagmolen, Smith, Smith, Smith, Somala, Son, Sonnenberg, Sorazu, Sorrentino, Souradeep, Spencer, Srivastava, Staats, Staley, Steinke, Steinlechner, Steinlechner, Steinmeyer, Stevenson, Stone, Stops, Strain, Stratta, Strigin, Strunk, Sturani, Stuver, Summerscales, Sun, Sunil, Suresh, Sutton, Swinkels, Szczepa\ifmmode~\acute{n}\else \'{n}\fi{}czyk, Tacca, Tait, Talbot, Talukder, Tanner, T\'apai, Taracchini, Tasson, Taylor, Taylor, Tewari, Theeg, Thies, Thomas, Thomas, Thomas, Thorne, Thorne, Thrane, Tiwari, Tiwari, Tokmakov, Toland, Tonelli, Tornasi, Torres-Forn\'e, Torrie, T\"oyr\"a, Travasso, Traylor, Trinastic, Tringali, Trozzo, Tsang, Tse, Tso, Tsukada, Tsuna, Tuyenbayev, Ueno, Ugolini, Unnikrishnan, Urban, Usman, Vahlbruch, Vajente, Valdes, Vallisneri, van Bakel, van Beuzekom, van~den Brand, Van Den~Broeck, Vander-Hyde, van~der Schaaf, van Heijningen, van Veggel, Vardaro, Varma, Vass, Vas\'uth, Vecchio, Vedovato, Veitch, Veitch, Venkateswara,
  Venugopalan, Verkindt, Vetrano, Vicer\'e, Viets, Vinciguerra, Vine, Vinet, Vitale, Vo, Vocca, Vorvick, Vyatchanin, Wade, Wade, Wade, Walet, Walker, Wallace, Walsh, Wang, Wang, Wang, Wang, Wang, Ward, Warner, Was, Watchi, Weaver, Wei, Weinert, Weinstein, Weiss, Wen, Wessel, We\ss{}els, Westerweck, Westphal, Wette, Whelan, Whitcomb, Whiting, Whittle, Wilken, Williams, Williams, Williamson, Willis, Willke, Wimmer, Winkler, Wipf, Wittel, Woan, Woehler, Wofford, Wong, Worden, Wright, Wu, Wysocki, Xiao, Yamamoto, Yancey, Yang, Yap, Yazback, Yu, Yu, Yvert, Zadro\ifmmode~\dot{z}\else \.{z}\fi{}ny, Zanolin, Zelenova, Zendri, Zevin, Zhang, Zhang, Zhang, Zhang, Zhao, Zhou, Zhou, Zhu, Zhu, Zimmerman, Zucker, \& Zweizig}]{PhysRevLett.119.161101}
Abbott, B.~P., Abbott, R., Abbott, T.~D., {et~al.} 2017, PhRvL, 119, 161101, \dodoi{10.1103/PhysRevLett.119.161101}

\bibitem[{{Abbott} {et~al.}(2023){Abbott}, {Abbott}, {Acernese}, {Ackley}, {Adams}, {Adhikari}, {Adhikari}, {Adya}, {Affeldt}, {Agarwal}, {Agathos}, {Agatsuma}, {Aggarwal}, {Aguiar}, {Aiello}, {Ain}, {Ajith}, {Akutsu}, {Albanesi}, {Allocca}, {Altin}, {Amato}, {Anand}, {Anand}, {Ananyeva}, {Anderson}, {Anderson}, {Ando}, {Andrade}, {Andres}, {Andri{\'c}}, {Angelova}, {Ansoldi}, {Antelis}, {Antier}, {Appert}, {Arai}, {Arai}, {Arai}, {Araki}, {Araya}, {Araya}, {Areeda}, {Ar{\`e}ne}, {Aritomi}, {Arnaud}, {Aronson}, {Arun}, {Asada}, {Asali}, {Ashton}, {Aso}, {Assiduo}, {Aston}, {Astone}, {Aubin}, {Austin}, {Babak}, {Badaracco}, {Bader}, {Badger}, {Bae}, {Bae}, {Baer}, {Bagnasco}, {Bai}, {Baiotti}, {Baird}, {Bajpai}, {Ball}, {Ballardin}, {Ballmer}, {Balsamo}, {Baltus}, {Banagiri}, {Bankar}, {Barayoga}, {Barbieri}, {Barish}, {Barker}, {Barneo}, {Barone}, {Barr}, {Barsotti}, {Barsuglia}, {Barta}, {Bartlett}, {Barton}, {Bartos}, {Bassiri}, {Basti}, {Bawaj}, {Bayley}, {Baylor}, {Bazzan}, {B{\'e}csy}, {Bedakihale},
  {Bejger}, {Belahcene}, {Benedetto}, {Beniwal}, {Bennett}, {Bentley}, {Benyaala}, {Bergamin}, {Berger}, {Bernuzzi}, {Berry}, {Bersanetti}, {Bertolini}, {Betzwieser}, {Beveridge}, {Bhandare}, {Bhardwaj}, {Bhattacharjee}, {Bhaumik}, {Bilenko}, {Billingsley}, {Bini}, {Birney}, {Birnholtz}, {Biscans}, {Bischi}, {Biscoveanu}, {Bisht}, {Biswas}, {Bitossi}, {Bizouard}, {Blackburn}, {Blair}, {Blair}, {Blair}, {Bobba}, {Bode}, {Boer}, {Bogaert}, {Boldrini}, {Bonavena}, {Bondu}, {Bonilla}, {Bonnand}, {Booker}, {Boom}, {Bork}, {Boschi}, {Bose}, {Bose}, {Bossilkov}, {Boudart}, {Bouffanais}, {Boumerdassi}, {Bozzi}, {Bradaschia}, {Brady}, {Bramley}, {Branch}, {Branchesi}, {Brau}, {Breschi}, {Briant}, {Briggs}, {Brillet}, {Brinkmann}, {Brockill}, {Brooks}, {Brooks}, {Brown}, {Brunett}, {Bruno}, {Bruntz}, {Bryant}, {Buchanan}, {Bulik}, {Bulten}, {Buonanno}, {Buscicchio}, {Buskulic}, {Buy}, {Byer}, {Cadonati}, {Cagnoli}, {Cahillane}, {Bustillo}, {Callaghan}, {Callister}, {Calloni}, {Cameron}, {Camp}, {Canepa}, {Canevarolo},
  {Cannavacciuolo}, {Cannon}, {Cao}, {Cao}, {Capocasa}, {Capote}, {Carapella}, {Carbognani}, {Carlin}, {Carney}, {Carpinelli}, {Carrillo}, {Carullo}, {Carver}, {Diaz}, {Casentini}, {Castaldi}, {Caudill}, {Cavagli{\`a}}, {Cavalier}, {Cavalieri}, {Ceasar}, {Cella}, {Cerd{\'a}-Dur{\'a}n}, {Cesarini}, {Chaibi}, {Chakravarti}, {Subrahmanya}, {Champion}, {Chan}, {Chan}, {Chan}, {Chan}, {Chan}, {Chandra}, {Chanial}, {Chao}, {Charlton}, {Chase}, {Chassande-Mottin}, {Chatterjee}, {Chatterjee}, {Chatterjee}, {Chaturvedi}, {Chaty}, {Chen}, {Chen}, {Chen}, {Chen}, {Chen}, {Chen}, {Chen}, {Chen}, {Cheng}, {Cheong}, {Cheung}, {Chia}, {Chiadini}, {Chiang}, {Chiarini}, {Chierici}, {Chincarini}, {Chiofalo}, {Chiummo}, {Cho}, {Cho}, {Choudhary}, {Choudhary}, {Christensen}, {Chu}, {Chu}, {Chu}, {Chua}, {Chung}, {Ciani}, {Ciecielag}, {Cie{\'s}lar}, {Cifaldi}, {Ciobanu}, {Ciolfi}, {Cipriano}, {Cirone}, {Clara}, {Clark}, {Clark}, {Clarke}, {Clearwater}, {Clesse}, {Cleva}, {Coccia}, {Codazzo}, {Cohadon}, {Cohen}, {Cohen},
  {Colleoni}, {Collette}, {Colombo}, {Colpi}, {Compton}, {Constancio}, {Conti}, {Cooper}, {Corban}, {Corbitt}, {Cordero-Carri{\'o}n}, {Corezzi}, {Corley}, {Cornish}, {Corre}, {Corsi}, {Cortese}, {Costa}, {Cotesta}, {Coughlin}, {Coulon}, {Countryman}, {Cousins}, {Couvares}, {Coward}, {Cowart}, {Coyne}, {Coyne}, {Creighton}, {Creighton}, {Criswell}, {Croquette}, {Crowder}, {Cudell}, {Cullen}, {Cumming}, {Cummings}, {Cunningham}, {Cuoco}, {Cury{\l}o}, {Dabadie}, {Dal Canton}, {Dall'Osso}, {D{\'a}lya}, {Dana}, {Daneshgaranbajastani}, {D'Angelo}, {Danilishin}, {D'Antonio}, {Danzmann}, {Darsow-Fromm}, {Dasgupta}, {Datrier}, {Datta}, {Dattilo}, {Dave}, {Davier}, {Davies}, {Davis}, {Davis}, {Daw}, {Dean}, {Debra}, {Deenadayalan}, {Degallaix}, {de Laurentis}, {Del{\'e}glise}, {Del Favero}, {de Lillo}, {de Lillo}, {Del Pozzo}, {Demarchi}, {de Matteis}, {D'Emilio}, {Demos}, {Dent}, {Depasse}, {de Pietri}, {De Rosa}, {de Rossi}, {Desalvo}, {de Simone}, {Dhurandhar}, {D{\'\i}az}, {Diaz-Ortiz}, {Didio}, {Dietrich}, {di
  Fiore}, {di Fronzo}, {di Giorgio}, {di Giovanni}, {di Giovanni}, {di Girolamo}, {di Lieto}, {Ding}, {di Pace}, {di Palma}, {di Renzo}, {Divakarla}, {Dmitriev}, {Doctor}, {D'Onofrio}, {Donovan}, {Dooley}, {Doravari}, {Dorrington}, {Drago}, {Driggers}, {Drori}, {Ducoin}, {Dupej}, {Durante}, {D'Urso}, {Duverne}, {Dwyer}, {Eassa}, {Easter}, {Ebersold}, {Eckhardt}, {Eddolls}, {Edelman}, {Edo}, {Edy}, {Effler}, {Eguchi}, {Eichholz}, {Eikenberry}, {Eisenmann}, {Eisenstein}, {Ejlli}, {Engelby}, {Enomoto}, {Errico}, {Essick}, {Estell{\'e}s}, {Estevez}, {Etienne}, {Etzel}, {Evans}, {Evans}, {Ewing}, {Fafone}, {Fair}, {Fairhurst}, {Farah}, {Farinon}, {Farr}, {Farr}, {Farrow}, {Fauchon-Jones}, {Favaro}, {Favata}, {Fays}, {Fazio}, {Feicht}, {Fejer}, {Fenyvesi}, {Ferguson}, {Fernandez-Galiana}, {Ferrante}, {Ferreira}, {Fidecaro}, {Figura}, {Fiori}, {Fishbach}, {Fisher}, {Fittipaldi}, {Fiumara}, {Flaminio}, {Floden}, {Fong}, {Font}, {Fornal}, {Forsyth}, {Franke}, {Frasca}, {Frasconi}, {Frederick}, {Freed}, {Frei},
  {Freise}, {Frey}, {Fritschel}, {Frolov}, {Fronz{\'e}}, {Fujii}, {Fujikawa}, {Fukunaga}, {Fukushima}, {Fulda}, {Fyffe}, {Gabbard}, {Gadre}, {Gair}, {Gais}, {Galaudage}, {Gamba}, {Ganapathy}, {Ganguly}, {Gao}, {Gaonkar}, {Garaventa}, {Garc{\'\i}a-N{\'u}{\~n}ez}, {Garc{\'\i}a-Quir{\'o}s}, {Garufi}, {Gateley}, {Gaudio}, {Gayathri}, {Ge}, {Gemme}, {Gennai}, {George}, {Gerberding}, {Gergely}, {Gewecke}, {Ghonge}, {Ghosh}, {Ghosh}, {Ghosh}, {Ghosh}, {Giacomazzo}, {Giacoppo}, {Giaime}, {Giardina}, {Gibson}, {Gier}, {Giesler}, {Giri}, {Gissi}, {Glanzer}, {Gleckl}, {Godwin}, {Goetz}, {Goetz}, {Gohlke}, {Goncharov}, {Gonz{\'a}lez}, {Gopakumar}, {Gosselin}, {Gouaty}, {Gould}, {Grace}, {Grado}, {Granata}, {Granata}, {Grant}, {Gras}, {Grassia}, {Gray}, {Gray}, {Greco}, {Green}, {Green}, {Gretarsson}, {Gretarsson}, {Griffith}, {Griffiths}, {Griggs}, {Grignani}, {Grimaldi}, {Grimm}, {Grote}, {Grunewald}, {Gruning}, {Guerra}, {Guidi}, {Guimaraes}, {Guix{\'e}}, {Gulati}, {Guo}, {Guo}, {Gupta}, {Gupta}, {Gupta}, {Gustafson},
  {Gustafson}, {Guzman}, {Ha}, {Haegel}, {Hagiwara}, {Haino}, {Halim}, {Hall}, {Hamilton}, {Hammond}, {Han}, {Haney}, {Hanks}, {Hanna}, {Hannam}, {Hannuksela}, {Hansen}, {Hansen}, {Hanson}, {Harder}, {Hardwick}, {Haris}, {Harms}, {Harry}, {Harry}, {Hartwig}, {Hasegawa}, {Haskell}, {Hasskew}, {Haster}, {Hattori}, {Haughian}, {Hayakawa}, {Hayama}, {Hayes}, {Healy}, {Heidmann}, {Heidt}, {Heintze}, {Heinze}, {Heinzel}, {Heitmann}, {Hellman}, {Hello}, {Helmling-Cornell}, {Hemming}, {Hendry}, {Heng}, {Hennes}, {Hennig}, {Hennig}, {Hernandez}, {Hernandez Vivanco}, {Heurs}, {Hild}, {Hill}, {Himemoto}, {Hines}, {Hiranuma}, {Hirata}, {Hirose}, {Hochheim}, {Hofman}, {Hohmann}, {Holcomb}, {Holland}, {Hollows}, {Holmes}, {Holt}, {Holz}, {Hong}, {Hopkins}, {Hough}, {Hourihane}, {Howell}, {Hoy}, {Hoyland}, {Hreibi}, {Hsieh}, {Hsu}, {Huang}, {Huang}, {Huang}, {Huang}, {Huang}, {Huang}, {H{\"u}bner}, {Huddart}, {Hughey}, {Hui}, {Hui}, {Husa}, {Huttner}, {Huxford}, {Huynh-Dinh}, {Ide}, {Idzkowski}, {Iess}, {Ikenoue}, {Imam},
  {Inayoshi}, {Ingram}, {Inoue}, {Ioka}, {Isi}, {Isleif}, {Ito}, {Itoh}, {Iyer}, {Izumi}, {Jaberianhamedan}, {Jacqmin}, {Jadhav}, {Jadhav}, {James}, {Jan}, {Jani}, {Janquart}, {Janssens}, {Janthalur}, {Jaranowski}, {Jariwala}, {Jaume}, {Jenkins}, {Jenner}, {Jeon}, {Jeunon}, {Jia}, {Jin}, {Johns}, {Jones}, {Jones}, {Jones}, {Jones}, {Jones}, {Jonker}, {Ju}, {Jung}, {Jung}, {Junker}, {Juste}, {Kaihotsu}, {Kajita}, {Kakizaki}, {Kalaghatgi}, {Kalogera}, {Kamai}, {Kamiizumi}, {Kanda}, {Kandhasamy}, {Kang}, {Kanner}, {Kao}, {Kapadia}, {Kapasi}, {Karat}, {Karathanasis}, {Karki}, {Kashyap}, {Kasprzack}, {Kastaun}, {Katsanevas}, {Katsavounidis}, {Katzman}, {Kaur}, {Kawabe}, {Kawaguchi}, {Kawai}, {Kawasaki}, {K{\'e}f{\'e}lian}, {Keitel}, {Key}, {Khadka}, {Khalili}, {Khan}, {Khazanov}, {Khetan}, {Khursheed}, {Kijbunchoo}, {Kim}, {Kim}, {Kim}, {Kim}, {Kim}, {Kim}, {Kimball}, {Kimura}, {Kinley-Hanlon}, {Kirchhoff}, {Kissel}, {Kita}, {Kitazawa}, {Kleybolte}, {Klimenko}, {Knee}, {Knowles}, {Knyazev}, {Koch}, {Koekoek},
  {Kojima}, {Kokeyama}, {Koley}, {Kolitsidou}, {Kolstein}, {Komori}, {Kondrashov}, {Kong}, {Kontos}, {Koper}, {Korobko}, {Kotake}, {Kovalam}, {Kozak}, {Kozakai}, {Kozu}, {Kringel}, {Krishnendu}, {Kr{\'o}lak}, {Kuehn}, {Kuei}, {Kuijer}, {Kumar}, {Kumar}, {Kumar}, {Kumar}, {Kume}, {Kuns}, {Kuo}, {Kuo}, {Kuromiya}, {Kuroyanagi}, {Kusayanagi}, {Kuwahara}, {Kwak}, {Lagabbe}, {Laghi}, {Lalande}, {Lam}, {Lamberts}, {Landry}, {Lane}, {Lang}, {Lange}, {Lantz}, {La Rosa}, {Lartaux-Vollard}, {Lasky}, {Laxen}, {Lazzarini}, {Lazzaro}, {Leaci}, {Leavey}, {Lecoeuche}, {Lee}, {Lee}, {Lee}, {Lee}, {Lee}, {Lee}, {Lehmann}, {Lema{\^\i}tre}, {Leonardi}, {Leroy}, {Letendre}, {Levesque}, {Levin}, {Leviton}, {Leyde}, {Li}, {Li}, {Li}, {Li}, {Li}, {Li}, {Lin}, {Lin}, {Lin}, {Lin}, {Lin}, {Linde}, {Linker}, {Linley}, {Littenberg}, {Liu}, {Liu}, {Liu}, {Liu}, {Llamas}, {Llorens-Monteagudo}, {Lo}, {Lockwood}, {London}, {Longo}, {Lopez}, {Lopez Portilla}, {Lorenzini}, {Loriette}, {Lormand}, {Losurdo}, {Lott}, {Lough}, {Lousto},
  {Lovelace}, {Lucaccioni}, {L{\"u}ck}, {Lumaca}, {Lundgren}, {Luo}, {Lynam}, {Macas}, {Macinnis}, {MacLeod}, {MacMillan}, {Macquet}, {Hernandez}, {Magazz{\`u}}, {Magee}, {Maggiore}, {Magnozzi}, {Mahesh}, {Majorana}, {Makarem}, {Maksimovic}, {Maliakal}, {Malik}, {Man}, {Mandic}, {Mangano}, {Mango}, {Mansell}, {Manske}, {Mantovani}, {Mapelli}, {Marchesoni}, {Marchio}, {Marion}, {Mark}, {M{\'a}rka}, {M{\'a}rka}, {Markakis}, {Markosyan}, {Markowitz}, {Maros}, {Marquina}, {Marsat}, {Martelli}, {Martin}, {Martin}, {Martinez}, {Martinez}, {Martinez}, {Martinovic}, {Martynov}, {Marx}, {Masalehdan}, {Mason}, {Massera}, {Masserot}, {Massinger}, {Masso-Reid}, {Mastrogiovanni}, {Matas}, {Mateu-Lucena}, {Matichard}, {Matiushechkina}, {Mavalvala}, {McCann}, {McCarthy}, {McClelland}, {McClincy}, {McCormick}, {McCuller}, {McGhee}, {McGuire}, {McIsaac}, {McIver}, {McRae}, {McWilliams}, {Meacher}, {Mehmet}, {Mehta}, {Meijer}, {Melatos}, {Melchor}, {Mendell}, {Menendez-Vazquez}, {Menoni}, {Mercer}, {Mereni}, {Merfeld},
  {Merilh}, {Merritt}, {Merzougui}, {Meshkov}, {Messenger}, {Messick}, {Meyers}, {Meylahn}, {Mhaske}, {Miani}, {Miao}, {Michaloliakos}, {Michel}, {Michimura}, {Middleton}, {Milano}, {Miller}, {Miller}, {Miller}, {Millhouse}, {Mills}, {Milotti}, {Minazzoli}, {Minenkov}, {Mio}, {Mir}, {Miravet-Ten{\'e}s}, {Mishra}, {Mishra}, {Mistry}, {Mitra}, {Mitrofanov}, {Mitselmakher}, {Mittleman}, {Miyakawa}, {Miyamoto}, {Miyazaki}, {Miyo}, {Miyoki}, {Mo}, {Moguel}, {Mogushi}, {Mohapatra}, {Mohite}, {Molina}, {Molina-Ruiz}, {Mondin}, {Montani}, {Moore}, {Moraru}, {Morawski}, {More}, {Moreno}, {Moreno}, {Mori}, {Morisaki}, {Moriwaki}, {Mours}, {Mow-Lowry}, {Mozzon}, {Muciaccia}, {Mukherjee}, {Mukherjee}, {Mukherjee}, {Mukherjee}, {Mukherjee}, {Mukund}, {Mullavey}, {Munch}, {Mu{\~n}iz}, {Murray}, {Musenich}, {Muusse}, {Nadji}, {Nagano}, {Nagano}, {Nagar}, {Nakamura}, {Nakano}, {Nakano}, {Nakashima}, {Nakayama}, {Napolano}, {Nardecchia}, {Narikawa}, {Naticchioni}, {Nayak}, {Nayak}, {Negishi}, {Neil}, {Neilson}, {Nelemans},
  {Nelson}, {Nery}, {Neubauer}, {Neunzert}, {Ng}, {Ng}, {Nguyen}, {Nguyen}, {Nguyen}, {Quynh}, {Ni}, {Nichols}, {Nishizawa}, {Nissanke}, {Nitoglia}, {Nocera}, {Norman}, {North}, {Nozaki}, {Nuttall}, {Oberling}, {O'Brien}, {Obuchi}, {O'Dell}, {Oelker}, {Ogaki}, {Oganesyan}, {Oh}, {Oh}, {Oh}, {Ohashi}, {Ohishi}, {Ohkawa}, {Ohme}, {Ohta}, {Okada}, {Okutani}, {Okutomi}, {Olivetto}, {Oohara}, {Ooi}, {Oram}, {O'Reilly}, {Ormiston}, {Ormsby}, {Ortega}, {O'Shaughnessy}, {O'Shea}, {Oshino}, {Ossokine}, {Osthelder}, {Otabe}, {Ottaway}, {Overmier}, {Pace}, {Pagano}, {Page}, {Pagliaroli}, {Pai}, {Pai}, {Palamos}, {Palashov}, {Palomba}, {Pan}, {Pan}, {Panda}, {Pang}, {Pang}, {Pankow}, {Pannarale}, {Pant}, {Panther}, {Paoletti}, {Paoli}, {Paolone}, {Parisi}, {Park}, {Park}, {Parker}, {Pascucci}, {Pasqualetti}, {Passaquieti}, {Passuello}, {Patel}, {Pathak}, {Patricelli}, {Patron}, {Patrone}, {Paul}, {Payne}, {Pedraza}, {Pegoraro}, {Pele}, {Arellano}, {Penn}, {Perego}, {Pereira}, {Pereira}, {Perez}, {P{\'e}rigois},
  {Perkins}, {Perreca}, {Perri{\`e}s}, {Petermann}, {Petterson}, {Pfeiffer}, {Pham}, {Phukon}, {Piccinni}, {Pichot}, {Piendibene}, {Piergiovanni}, {Pierini}, {Pierro}, {Pillant}, {Pillas}, {Pilo}, {Pinard}, {Pinto}, {Pinto}, {Piotrzkowski}, {Piotrzkowski}, {Pirello}, {Pitkin}, {Placidi}, {Planas}, {Plastino}, {Pluchar}, {Poggiani}, {Polini}, {Pong}, {Ponrathnam}, {Popolizio}, {Porter}, {Poulton}, {Powell}, {Pracchia}, {Pradier}, {Prajapati}, {Prasai}, {Prasanna}, {Pratten}, {Principe}, {Prodi}, {Prokhorov}, {Prosposito}, {Prudenzi}, {Puecher}, {Punturo}, {Puosi}, {Puppo}, {P{\"u}rrer}, {Qi}, {Quetschke}, {Quitzow-James}, {Raab}, {Raaijmakers}, {Radkins}, {Radulesco}, {Raffai}, {Rail}, {Raja}, {Rajan}, {Ramirez}, {Ramirez}, {Ramos-Buades}, {Rana}, {Rapagnani}, {Rapol}, {Ray}, {Raymond}, {Raza}, {Razzano}, {Read}, {Rees}, {Regimbau}, {Rei}, {Reid}, {Reid}, {Reitze}, {Relton}, {Renzini}, {Rettegno}, {Rezac}, {Ricci}, {Richards}, {Richardson}, {Richardson}, {Riemenschneider}, {Riles}, {Rinaldi}, {Rink}, {Rizzo},
  {Robertson}, {Robie}, {Robinet}, {Rocchi}, {Rodriguez}, {Rolland}, {Rollins}, {Romanelli}, {Romano}, {Romel}, {Romero-Rodr{\'\i}guez}, {Romero-Shaw}, {Romie}, {Ronchini}, {Rosa}, {Rose}, {Rosi{\'n}ska}, {Ross}, {Rowan}, {Rowlinson}, {Roy}, {Roy}, {Roy}, {Rozza}, {Ruggi}, {Ryan}, {Sachdev}, {Sadecki}, {Sadiq}, {Sago}, {Saito}, {Saito}, {Sakai}, {Sakai}, {Sakellariadou}, {Sakuno}, {Salafia}, {Salconi}, {Saleem}, {Salemi}, {Samajdar}, {Sanchez}, {Sanchez}, {Sanchez}, {Sanchis-Gual}, {Sanders}, {Sanuy}, {Saravanan}, {Sarin}, {Sassolas}, {Satari}, {Sato}, {Sato}, {Sauter}, {Savage}, {Sawada}, {Sawant}, {Sawant}, {Sayah}, {Schaetzl}, {Scheel}, {Scheuer}, {Schiworski}, {Schmidt}, {Schmidt}, {Schnabel}, {Schneewind}, {Schofield}, {Sch{\"o}nbeck}, {Schulte}, {Schutz}, {Schwartz}, {Scott}, {Scott}, {Seglar-Arroyo}, {Sekiguchi}, {Sekiguchi}, {Sellers}, {Sengupta}, {Sentenac}, {Seo}, {Sequino}, {Sergeev}, {Setyawati}, {Shaffer}, {Shahriar}, {Shams}, {Shao}, {Sharma}, {Sharma}, {Shawhan}, {Shcheblanov}, {Shibagaki},
  {Shikauchi}, {Shimizu}, {Shimoda}, {Shimode}, {Shinkai}, {Shishido}, {Shoda}, {Shoemaker}, {Shoemaker}, {Shyamsundar}, {Sieniawska}, {Sigg}, {Singer}, {Singh}, {Singh}, {Singha}, {Sintes}, {Sipala}, {Skliris}, {Slagmolen}, {Slaven-Blair}, {Smetana}, {Smith}, {Smith}, {Soldateschi}, {Somala}, {Somiya}, {Son}, {Soni}, {Soni}, {Sordini}, {Sorrentino}, {Sorrentino}, {Sotani}, {Soulard}, {Souradeep}, {Sowell}, {Spagnuolo}, {Spencer}, {Spera}, {Srinivasan}, {Srivastava}, {Srivastava}, {Staats}, {Stachie}, {Steer}, {Steinlechner}, {Steinlechner}, {Stops}, {Stover}, {Strain}, {Strang}, {Stratta}, {Strunk}, {Sturani}, {Stuver}, {Sudhagar}, {Sudhir}, {Sugimoto}, {Suh}, {Summerscales}, {Sun}, {Sun}, {Sunil}, {Sur}, {Suresh}, {Sutton}, {Suzuki}, {Suzuki}, {Swinkels}, {Szczepa{\'n}czyk}, {Szewczyk}, {Tacca}, {Tagoshi}, {Tait}, {Takahashi}, {Takahashi}, {Takamori}, {Takano}, {Takeda}, {Takeda}, {Talbot}, {Talbot}, {Tanaka}, {Tanaka}, {Tanaka}, {Tanaka}, {Tanaka}, {Tanasijczuk}, {Tanioka}, {Tanner}, {Tao}, {Tao}, {Tapia
  San Martin}, {Tapia San Mart{\'\i}n}, {Taranto}, {Tasson}, {Telada}, {Tenorio}, {Terhune}, {Terkowski}, {Thirugnanasambandam}, {Thomas}, {Thomas}, {Thompson}, {Thondapu}, {Thorne}, {Thrane}, {Tiwari}, {Tiwari}, {Tiwari}, {Toivonen}, {Toland}, {Tolley}, {Tomaru}, {Tomigami}, {Tomura}, {Tonelli}, {Torres-Forn{\'e}}, {Torrie}, {E Melo}, {T{\"o}yr{\"a}}, {Trapananti}, {Travasso}, {Traylor}, {Trevor}, {Tringali}, {Tripathee}, {Troiano}, {Trovato}, {Trozzo}, {Trudeau}, {Tsai}, {Tsai}, {Tsang}, {Tsang}, {Tsao}, {Tse}, {Tso}, {Tsubono}, {Tsuchida}, {Tsukada}, {Tsuna}, {Tsutsui}, {Tsuzuki}, {Turbang}, {Turconi}, {Tuyenbayev}, {Ubhi}, {Uchikata}, {Uchiyama}, {Udall}, {Ueda}, {Uehara}, {Ueno}, {Ueshima}, {Unnikrishnan}, {Uraguchi}, {Urban}, {Ushiba}, {Utina}, {Vahlbruch}, {Vajente}, {Vajpeyi}, {Valdes}, {Valentini}, {Valsan}, {van Bakel}, {van Beuzekom}, {van den Brand}, {van den Broeck}, {Vander-Hyde}, {van der Schaaf}, {van Heijningen}, {Vanosky}, {van Putten}, {van Remortel}, {Vardaro}, {Vargas}, {Varma},
  {Vas{\'u}th}, {Vecchio}, {Vedovato}, {Veitch}, {Veitch}, {Venneberg}, {Venugopalan}, {Verkindt}, {Verma}, {Verma}, {Veske}, {Vetrano}, {Vicer{\'e}}, {Vidyant}, {Viets}, {Vijaykumar}, {Villa-Ortega}, {Vinet}, {Virtuoso}, {Vitale}, {Vo}, {Vocca}, {von Reis}, {von Wrangel}, {Vorvick}, {Vyatchanin}, {Wade}, {Wade}, {Wagner}, {Walet}, {Walker}, {Wallace}, {Wallace}, {Walsh}, {Wang}, {Wang}, {Wang}, {Ward}, {Warner}, {Was}, {Washimi}, {Washington}, {Watada}, {Watchi}, {Weaver}, {Webster}, {Weinert}, {Weinstein}, {Weiss}, {Weller}, {Wellmann}, {Wen}, {We{\ss}els}, {Wette}, {Whelan}, {White}, {Whiting}, {Whittle}, {Wilken}, {Williams}, {Williams}, {Williamson}, {Willis}, {Willke}, {Wilson}, {Winkler}, {Wipf}, {Wlodarczyk}, {Woan}, {Woehler}, {Wofford}, {Wong}, {Wu}, {Wu}, {Wu}, {Wu}, {Wysocki}, {Xiao}, {Xu}, {Yamada}, {Yamamoto}, {Yamamoto}, {Yamamoto}, {Yamamoto}, {Yamashita}, {Yamazaki}, {Yang}, {Yang}, {Yang}, {Yang}, {Yang}, {Yap}, {Yeeles}, {Yelikar}, {Ying}, {Yokogawa}, {Yokoyama}, {Yokozawa}, {Yoo},
  {Yoshioka}, {Yu}, {Yu}, {Yuzurihara}, {Zadro{\.z}ny}, {Zanolin}, {Zeidler}, {Zelenova}, {Zendri}, {Zevin}, {Zhan}, {Zhang}, {Zhang}, {Zhang}, {Zhang}, {Zhang}, {Zhao}, {Zhao}, {Zhao}, {Zhao}, {Zhou}, {Zhou}, {Zhu}, {Zhu}, {Zimmerman}, {Zucker}, {Zweizig}, {Bhardwaj}, {Boyle}, {Cassanelli}, {Dong}, {Fonseca}, {Kaspi}, {Leung}, {Masui}, {Meyers}, {Michilli}, {Ng}, {Pearlman}, {Petroff}, {Pleunis}, {Rafiei-Ravandi}, {Rahman}, {Ransom}, {Scholz}, {Shin}, {Smith}, {Stairs}, {Tendulkar}, {Zwaniga}, {Ligo Scientific Collaboration}, {VIRGO Collaboration}, {Kagra Collaboration}, \& {Chime/Frb Collaboration}}]{2023ApJ...955..155A}
{Abbott}, R., {Abbott}, T.~D., {Acernese}, F., {et~al.} 2023, \apj, 955, 155, \dodoi{10.3847/1538-4357/acd770}

\bibitem[{{Anna-Thomas} {et~al.}(2023){Anna-Thomas}, {Connor}, {Dai}, {Feng}, {Burke-Spolaor}, {Beniamini}, {Yang}, {Zhang}, {Aggarwal}, {Law}, {Li}, {Niu}, {Chatterjee}, {Cruces}, {Duan}, {Filipovic}, {Hobbs}, {Lynch}, {Miao}, {Niu}, {Ocker}, {Tsai}, {Wang}, {Xue}, {Yao}, {Yu}, {Zhang}, {Zhang}, {Zhu}, \& {Zhu}}]{2023Sci...380..599A}
{Anna-Thomas}, R., {Connor}, L., {Dai}, S., {et~al.} 2023, Science, 380, 599, \dodoi{10.1126/science.abo6526}

\bibitem[{Ashton {et~al.}(2019)}]{bilby_paper}
Ashton, G., {et~al.} 2019, ApJS, 241, 27, \dodoi{10.3847/1538-4365/ab06fc}

\bibitem[{{Bannister} {et~al.}(2019){Bannister}, {Deller}, {Phillips}, {Macquart}, {Prochaska}, {Tejos}, {Ryder}, {Sadler}, {Shannon}, {Simha}, {Day}, {McQuinn}, {North-Hickey}, {Bhandari}, {Arcus}, {Bennert}, {Burchett}, {Bouwhuis}, {Dodson}, {Ekers}, {Farah}, {Flynn}, {James}, {Kerr}, {Lenc}, {Mahony}, {O'Meara}, {Os{\l}owski}, {Qiu}, {Treu}, {U}, {Bateman}, {Bock}, {Bolton}, {Brown}, {Bunton}, {Chippendale}, {Cooray}, {Cornwell}, {Gupta}, {Hayman}, {Kesteven}, {Koribalski}, {MacLeod}, {McClure-Griffiths}, {Neuhold}, {Norris}, {Pilawa}, {Qiao}, {Reynolds}, {Roxby}, {Shimwell}, {Voronkov}, \& {Wilson}}]{2019Sci...365..565B}
{Bannister}, K.~W., {Deller}, A.~T., {Phillips}, C., {et~al.} 2019, Science, 365, 565, \dodoi{10.1126/science.aaw5903}

\bibitem[{{Bhandari} {et~al.}(2020){Bhandari}, {Sadler}, {Prochaska}, {Simha}, {Ryder}, {Marnoch}, {Bannister}, {Macquart}, {Flynn}, {Shannon}, {Tejos}, {Corro-Guerra}, {Day}, {Deller}, {Ekers}, {Lopez}, {Mahony}, {Nu{\~n}ez}, \& {Phillips}}]{2020ApJ...895L..37B}
{Bhandari}, S., {Sadler}, E.~M., {Prochaska}, J.~X., {et~al.} 2020, \apjl, 895, L37, \dodoi{10.3847/2041-8213/ab672e}

\bibitem[{{Bhandari} {et~al.}(2022){Bhandari}, {Heintz}, {Aggarwal}, {Marnoch}, {Day}, {Sydnor}, {Burke-Spolaor}, {Law}, {Xavier Prochaska}, {Tejos}, {Bannister}, {Butler}, {Deller}, {Ekers}, {Flynn}, {Fong}, {James}, {Lazio}, {Luo}, {Mahony}, {Ryder}, {Sadler}, {Shannon}, {Han}, {Lee}, \& {Zhang}}]{2022AJ....163...69B}
{Bhandari}, S., {Heintz}, K.~E., {Aggarwal}, K., {et~al.} 2022, \aj, 163, 69, \dodoi{10.3847/1538-3881/ac3aec}

\bibitem[{{Bhandari} {et~al.}(2023){Bhandari}, {Gordon}, {Scott}, {Marnoch}, {Sridhar}, {Kumar}, {James}, {Qiu}, {Bannister}, {T. Deller}, {Eftekhari}, {Fong}, {Glowacki}, {Prochaska}, {Ryder}, {Shannon}, \& {Simha}}]{2023ApJ...948...67B}
{Bhandari}, S., {Gordon}, A.~C., {Scott}, D.~R., {et~al.} 2023, \apj, 948, 67, \dodoi{10.3847/1538-4357/acc178}

\bibitem[{{Bhardwaj} {et~al.}(2021{\natexlab{a}}){Bhardwaj}, {Gaensler}, {Kaspi}, {Landecker}, {Mckinven}, {Michilli}, {Pleunis}, {Tendulkar}, {Andersen}, {Boyle}, {Cassanelli}, {Chawla}, {Cook}, {Dobbs}, {Fonseca}, {Kaczmarek}, {Leung}, {Masui}, {Mnchmeyer}, {Ng}, {Rafiei-Ravandi}, {Scholz}, {Shin}, {Smith}, {Stairs}, \& {Zwaniga}}]{2021ApJ...910L..18B}
{Bhardwaj}, M., {Gaensler}, B.~M., {Kaspi}, V.~M., {et~al.} 2021{\natexlab{a}}, \apjl, 910, L18, \dodoi{10.3847/2041-8213/abeaa6}

\bibitem[{{Bhardwaj} {et~al.}(2021{\natexlab{b}}){Bhardwaj}, {Kirichenko}, {Michilli}, {Mayya}, {Kaspi}, {Gaensler}, {Rahman}, {Tendulkar}, {Fonseca}, {Josephy}, {Leung}, {Merryfield}, {Petroff}, {Pleunis}, {Sanghavi}, {Scholz}, {Shin}, {Smith}, \& {Stairs}}]{2021ApJ...919L..24B}
{Bhardwaj}, M., {Kirichenko}, A.~Y., {Michilli}, D., {et~al.} 2021{\natexlab{b}}, \apjl, 919, L24, \dodoi{10.3847/2041-8213/ac223b}

\bibitem[{{Bhardwaj} {et~al.}(2024){Bhardwaj}, {Michilli}, {Kirichenko}, {Modilim}, {Shin}, {Kaspi}, {Andersen}, {Cassanelli}, {Brar}, {Chatterjee}, {Cook}, {Dong}, {Fonseca}, {Gaensler}, {Ibik}, {Kaczmarek}, {Lanman}, {Leung}, {Masui}, {Pandhi}, {Pearlman}, {Petroff}, {Pleunis}, {Prochaska}, {Rafiei-Ravandi}, {Sand}, {Scholz}, \& {Smith}}]{2024ApJ...971L..51B}
{Bhardwaj}, M., {Michilli}, D., {Kirichenko}, A.~Y., {et~al.} 2024, \apjl, 971, L51, \dodoi{10.3847/2041-8213/ad64d1}

\bibitem[{{Bochenek} {et~al.}(2020){Bochenek}, {Ravi}, {Belov}, {Hallinan}, {Kocz}, {Kulkarni}, \& {McKenna}}]{2020Natur.587...59B}
{Bochenek}, C.~D., {Ravi}, V., {Belov}, K.~V., {et~al.} 2020, \nat, 587, 59, \dodoi{10.1038/s41586-020-2872-x}

\bibitem[{{Bochenek} {et~al.}(2021){Bochenek}, {Ravi}, \& {Dong}}]{2021ApJ...907L..31B}
{Bochenek}, C.~D., {Ravi}, V., \& {Dong}, D. 2021, \apjl, 907, L31, \dodoi{10.3847/2041-8213/abd634}

\bibitem[{{Botticella} {et~al.}(2008){Botticella}, {Riello}, {Cappellaro}, {Benetti}, {Altavilla}, {Pastorello}, {Turatto}, {Greggio}, {Patat}, {Valenti}, {Zampieri}, {Harutyunyan}, {Pignata}, \& {Taubenberger}}]{2008A&A...479...49B}
{Botticella}, M.~T., {Riello}, M., {Cappellaro}, E., {et~al.} 2008, \aap, 479, 49, \dodoi{10.1051/0004-6361:20078011}

\bibitem[{{Caleb} {et~al.}(2023){Caleb}, {Driessen}, {Gordon}, {Tejos}, {Bernales}, {Qiu}, {Chibueze}, {Stappers}, {Rajwade}, {Cavallaro}, {Wang}, {Kumar}, {Majid}, {Wharton}, {Naudet}, {Bezuidenhout}, {Jankowski}, {Malenta}, {Morello}, {Sanidas}, {Surnis}, {Barr}, {Chen}, {Kramer}, {Fong}, {Kilpatrick}, {Prochaska}, {Simha}, {Venter}, {Heywood}, {Kundu}, \& {Schussler}}]{2023MNRAS.524.2064C}
{Caleb}, M., {Driessen}, L.~N., {Gordon}, A.~C., {et~al.} 2023, \mnras, 524, 2064, \dodoi{10.1093/mnras/stad1839}

\bibitem[{{Cassanelli} {et~al.}(2024){Cassanelli}, {Leung}, {Sanghavi}, {Mena-Parra}, {Cary}, {Mckinven}, {Bhardwaj}, {Masui}, {Michilli}, {Bandura}, {Chatterjee}, {Peterson}, {Kaczmarek}, {Rahman}, {Shin}, {Vanderlinde}, {Berger}, {Brar}, {Boyle}, {Breitman}, {Chawla}, {Curtin}, {Dobbs}, {Dong}, {Fonseca}, {Gaensler}, {Ibik}, {Kaspi}, {Khairy}, {Lanman}, {Lazda}, {Lin}, {Luo}, {Meyers}, {Milutinovic}, {Ng}, {Noble}, {Pearlman}, {Pen}, {Petroff}, {Pleunis}, {Quine}, {Rafiei-Ravandi}, {Renard}, {Sand}, {Schoen}, {Scholz}, {Smith}, {Stairs}, \& {Tendulkar}}]{2024NatAs...8.1429C}
{Cassanelli}, T., {Leung}, C., {Sanghavi}, P., {et~al.} 2024, Nature Astronomy, 8, 1429, \dodoi{10.1038/s41550-024-02357-x}

\bibitem[{{Chapman} {et~al.}(2007){Chapman}, {Tanvir}, {Priddey}, \& {Levan}}]{2007MNRAS.382L..21C}
{Chapman}, R., {Tanvir}, N.~R., {Priddey}, R.~S., \& {Levan}, A.~J. 2007, \mnras, 382, L21, \dodoi{10.1111/j.1745-3933.2007.00381.x}

\bibitem[{{Chen} {et~al.}(2024){Chen}, {Jia}, {Dong}, \& {Wang}}]{2024ApJ...973L..54C}
{Chen}, J.~H., {Jia}, X.~D., {Dong}, X.~F., \& {Wang}, F.~Y. 2024, \apjl, 973, L54, \dodoi{10.3847/2041-8213/ad7b39}

\bibitem[{{CHIME Collaboration} {et~al.}(2022){CHIME Collaboration}, {Amiri}, {Bandura}, {Boskovic}, {Chen}, {Cliche}, {Deng}, {Denman}, {Dobbs}, {Fandino}, {Foreman}, {Halpern}, {Hanna}, {Hill}, {Hinshaw}, {H{\"o}fer}, {Kania}, {Klages}, {Landecker}, {MacEachern}, {Masui}, {Mena-Parra}, {Milutinovic}, {Mirhosseini}, {Newburgh}, {Nitsche}, {Ordog}, {Pen}, {Pinsonneault-Marotte}, {Polzin}, {Reda}, {Renard}, {Shaw}, {Siegel}, {Singh}, {Smegal}, {Tretyakov}, {van Gassen}, {Vanderlinde}, {Wang}, {Wiebe}, {Willis}, \& {Wulf}}]{2022ApJS..261...29C}
{CHIME Collaboration}, {Amiri}, M., {Bandura}, K., {et~al.} 2022, \apjs, 261, 29, \dodoi{10.3847/1538-4365/ac6fd9}

\bibitem[{{CHIME/FRB Collaboration} {et~al.}(2018){CHIME/FRB Collaboration}, {Amiri}, {Bandura}, {Berger}, {Bhardwaj}, {Boyce}, {Boyle}, {Brar}, {Burhanpurkar}, {Chawla}, {Chowdhury}, {Cliche}, {Cranmer}, {Cubranic}, {Deng}, {Denman}, {Dobbs}, {Fandino}, {Fonseca}, {Gaensler}, {Giri}, {Gilbert}, {Good}, {Guliani}, {Halpern}, {Hinshaw}, {H{\"o}fer}, {Josephy}, {Kaspi}, {Landecker}, {Lang}, {Liao}, {Masui}, {Mena-Parra}, {Naidu}, {Newburgh}, {Ng}, {Patel}, {Pen}, {Pinsonneault-Marotte}, {Pleunis}, {Rafiei Ravandi}, {Ransom}, {Renard}, {Scholz}, {Sigurdson}, {Siegel}, {Smith}, {Stairs}, {Tendulkar}, {Vanderlinde}, \& {Wiebe}}]{2018ApJ...863...48C}
{CHIME/FRB Collaboration}, {Amiri}, M., {Bandura}, K., {et~al.} 2018, \apj, 863, 48, \dodoi{10.3847/1538-4357/aad188}

\bibitem[{{CHIME/FRB Collaboration} {et~al.}(2019{\natexlab{a}}){CHIME/FRB Collaboration}, {Amiri}, {Bandura}, {Bhardwaj}, {Boubel}, {Boyce}, {Boyle}, {Brar}, {Burhanpurkar}, {Chawla}, {Cliche}, {Cubranic}, {Deng}, {Denman}, {Dobbs}, {Fandino}, {Fonseca}, {Gaensler}, {Gilbert}, {Giri}, {Good}, {Halpern}, {Hanna}, {Hill}, {Hinshaw}, {H{\"o}fer}, {Josephy}, {Kaspi}, {Landecker}, {Lang}, {Masui}, {Mckinven}, {Mena-Parra}, {Merryfield}, {Milutinovic}, {Moatti}, {Naidu}, {Newburgh}, {Ng}, {Patel}, {Pen}, {Pinsonneault-Marotte}, {Pleunis}, {Rafiei-Ravandi}, {Ransom}, {Renard}, {Scholz}, {Shaw}, {Siegel}, {Smith}, {Stairs}, {Tendulkar}, {Tretyakov}, {Vanderlinde}, \& {Yadav}}]{2019Natur.566..230C}
---. 2019{\natexlab{a}}, \nat, 566, 230, \dodoi{10.1038/s41586-018-0867-7}

\bibitem[{{CHIME/FRB Collaboration} {et~al.}(2019{\natexlab{b}}){CHIME/FRB Collaboration}, {Amiri}, {Bandura}, {Bhardwaj}, {Boubel}, {Boyce}, {Boyle}, {. Brar}, {Burhanpurkar}, {Cassanelli}, {Chawla}, {Cliche}, {Cubranic}, {Deng}, {Denman}, {Dobbs}, {Fandino}, {Fonseca}, {Gaensler}, {Gilbert}, {Gill}, {Giri}, {Good}, {Halpern}, {Hanna}, {Hill}, {Hinshaw}, {H{\"o}fer}, {Josephy}, {Kaspi}, {Landecker}, {Lang}, {Lin}, {Masui}, {Mckinven}, {Mena-Parra}, {Merryfield}, {Michilli}, {Milutinovic}, {Moatti}, {Naidu}, {Newburgh}, {Ng}, {Patel}, {Pen}, {Pinsonneault-Marotte}, {Pleunis}, {Rafiei-Ravandi}, {Rahman}, {Ransom}, {Renard}, {Scholz}, {Shaw}, {Siegel}, {Smith}, {Stairs}, {Tendulkar}, {Tretyakov}, {Vanderlinde}, \& {Yadav}}]{2019Natur.566..235C}
---. 2019{\natexlab{b}}, \nat, 566, 235, \dodoi{10.1038/s41586-018-0864-x}

\bibitem[{{CHIME/FRB Collaboration} {et~al.}(2019{\natexlab{c}}){CHIME/FRB Collaboration}, {Andersen}, {Bandura}, {Bhardwaj}, {Boubel}, {Boyce}, {Boyle}, {Brar}, {Cassanelli}, {Chawla}, {Cubranic}, {Deng}, {Dobbs}, {Fandino}, {Fonseca}, {Gaensler}, {Gilbert}, {Giri}, {Good}, {Halpern}, {Hill}, {Hinshaw}, {H{\"o}fer}, {Josephy}, {Kaspi}, {Kothes}, {Landecker}, {Lang}, {Li}, {Lin}, {Masui}, {Mena-Parra}, {Merryfield}, {Mckinven}, {Michilli}, {Milutinovic}, {Naidu}, {Newburgh}, {Ng}, {Patel}, {Pen}, {Pinsonneault-Marotte}, {Pleunis}, {Rafiei-Ravandi}, {Rahman}, {Ransom}, {Renard}, {Scholz}, {Siegel}, {Singh}, {Smith}, {Stairs}, {Tendulkar}, {Tretyakov}, {Vanderlinde}, {Yadav}, \& {Zwaniga}}]{2019ApJ...885L..24C}
{CHIME/FRB Collaboration}, {Andersen}, B.~C., {Bandura}, K., {et~al.} 2019{\natexlab{c}}, \apjl, 885, L24, \dodoi{10.3847/2041-8213/ab4a80}

\bibitem[{{CHIME/FRB Collaboration} {et~al.}(2020{\natexlab{a}}){CHIME/FRB Collaboration}, {Andersen}, {Bandura}, {Bhardwaj}, {Bij}, {Boyce}, {Boyle}, {Brar}, {Cassanelli}, {Chawla}, {Chen}, {Cliche}, {Cook}, {Cubranic}, {Curtin}, {Denman}, {Dobbs}, {Dong}, {Fandino}, {Fonseca}, {Gaensler}, {Giri}, {Good}, {Halpern}, {Hill}, {Hinshaw}, {H{\"o}fer}, {Josephy}, {Kania}, {Kaspi}, {Landecker}, {Leung}, {Li}, {Lin}, {Masui}, {McKinven}, {Mena-Parra}, {Merryfield}, {Meyers}, {Michilli}, {Milutinovic}, {Mirhosseini}, {M{\"u}nchmeyer}, {Naidu}, {Newburgh}, {Ng}, {Patel}, {Pen}, {Pinsonneault-Marotte}, {Pleunis}, {Quine}, {Rafiei-Ravandi}, {Rahman}, {Ransom}, {Renard}, {Sanghavi}, {Scholz}, {Shaw}, {Shin}, {Siegel}, {Singh}, {Smegal}, {Smith}, {Stairs}, {Tan}, {Tendulkar}, {Tretyakov}, {Vanderlinde}, {Wang}, {Wulf}, \& {Zwaniga}}]{2020Natur.587...54C}
{CHIME/FRB Collaboration}, {Andersen}, B.~C., {Bandura}, K.~M., {et~al.} 2020{\natexlab{a}}, \nat, 587, 54, \dodoi{10.1038/s41586-020-2863-y}

\bibitem[{{CHIME/FRB Collaboration} {et~al.}(2020{\natexlab{b}}){CHIME/FRB Collaboration}, {Amiri}, {Andersen}, {Bandura}, {Bhardwaj}, {Boyle}, {Brar}, {Chawla}, {Chen}, {Cliche}, {Cubranic}, {Deng}, {Denman}, {Dobbs}, {Dong}, {Fandino}, {Fonseca}, {Gaensler}, {Giri}, {Good}, {Halpern}, {Hessels}, {Hill}, {H{\"o}fer}, {Josephy}, {Kania}, {Karuppusamy}, {Kaspi}, {Keimpema}, {Kirsten}, {Landecker}, {Lang}, {Leung}, {Li}, {Lin}, {Marcote}, {Masui}, {McKinven}, {Mena-Parra}, {Merryfield}, {Michilli}, {Milutinovic}, {Mirhosseini}, {Naidu}, {Newburgh}, {Ng}, {Nimmo}, {Paragi}, {Patel}, {Pen}, {Pinsonneault-Marotte}, {Pleunis}, {Rafiei-Ravandi}, {Rahman}, {Ransom}, {Renard}, {Sanghavi}, {Scholz}, {Shaw}, {Shin}, {Siegel}, {Singh}, {Smegal}, {Smith}, {Stairs}, {Tendulkar}, {Tretyakov}, {Vanderlinde}, {Wang}, {Wang}, {Wulf}, {Yadav}, \& {Zwaniga}}]{2020Natur.582..351C}
{CHIME/FRB Collaboration}, {Amiri}, M., {Andersen}, B.~C., {et~al.} 2020{\natexlab{b}}, \nat, 582, 351, \dodoi{10.1038/s41586-020-2398-2}

\bibitem[{{CHIME/FRB Collaboration} {et~al.}(2021){CHIME/FRB Collaboration}, {Amiri}, {Andersen}, {Bandura}, {Berger}, {Bhardwaj}, {Boyce}, {Boyle}, {Brar}, {Breitman}, {Cassanelli}, {Chawla}, {Chen}, {Cliche}, {Cook}, {Cubranic}, {Curtin}, {Deng}, {Dobbs}, {Dong}, {Eadie}, {Fandino}, {Fonseca}, {Gaensler}, {Giri}, {Good}, {Halpern}, {Hill}, {Hinshaw}, {Josephy}, {Kaczmarek}, {Kader}, {Kania}, {Kaspi}, {Landecker}, {Lang}, {Leung}, {Li}, {Lin}, {Masui}, {McKinven}, {Mena-Parra}, {Merryfield}, {Meyers}, {Michilli}, {Milutinovic}, {Mirhosseini}, {M{\"u}nchmeyer}, {Naidu}, {Newburgh}, {Ng}, {Patel}, {Pen}, {Petroff}, {Pinsonneault-Marotte}, {Pleunis}, {Rafiei-Ravandi}, {Rahman}, {Ransom}, {Renard}, {Sanghavi}, {Scholz}, {Shaw}, {Shin}, {Siegel}, {Sikora}, {Singh}, {Smith}, {Stairs}, {Tan}, {Tendulkar}, {Vanderlinde}, {Wang}, {Wulf}, \& {Zwaniga}}]{2021ApJS..257...59C}
---. 2021, \apjs, 257, 59, \dodoi{10.3847/1538-4365/ac33ab}

\bibitem[{{CHIME/FRB Collaboration} {et~al.}(2023){CHIME/FRB Collaboration}, {Andersen}, {Bandura}, {Bhardwaj}, {Boyle}, {Brar}, {Cassanelli}, {Chatterjee}, {Chawla}, {Cook}, {Curtin}, {Dobbs}, {Dong}, {Faber}, {Fandino}, {Fonseca}, {Gaensler}, {Giri}, {Herrera-Martin}, {Hill}, {Ibik}, {Josephy}, {Kaczmarek}, {Kader}, {Kaspi}, {Landecker}, {Lanman}, {Lazda}, {Leung}, {Lin}, {Masui}, {McKinven}, {Mena-Parra}, {Meyers}, {Michilli}, {Ng}, {Pandhi}, {Pearlman}, {Pen}, {Petroff}, {Pleunis}, {Rafiei-Ravandi}, {Rahman}, {Ransom}, {Renard}, {Sand}, {Sanghavi}, {Scholz}, {Shah}, {Shin}, {Siegel}, {Smith}, {Stairs}, {Su}, {Tendulkar}, {Vanderlinde}, {Wang}, {Wulf}, \& {Zwaniga}}]{2023ApJ...947...83C}
{CHIME/FRB Collaboration}, {Andersen}, B.~C., {Bandura}, K., {et~al.} 2023, \apj, 947, 83, \dodoi{10.3847/1538-4357/acc6c1}

\bibitem[{{Cordes} \& {Lazio}(2002)}]{2002astro.ph..7156C}
{Cordes}, J.~M., \& {Lazio}, T.~J.~W. 2002, arXiv e-prints, astro, \dodoi{10.48550/arXiv.astro-ph/0207156}

\bibitem[{{Coward} {et~al.}(2012){Coward}, {Howell}, {Piran}, {Stratta}, {Branchesi}, {Bromberg}, {Gendre}, {Burman}, \& {Guetta}}]{2012MNRAS.425.2668C}
{Coward}, D.~M., {Howell}, E.~J., {Piran}, T., {et~al.} 2012, \mnras, 425, 2668, \dodoi{10.1111/j.1365-2966.2012.21604.x}

\bibitem[{{Dilday} {et~al.}(2010){Dilday}, {Smith}, {Bassett}, {Becker}, {Bender}, {Castander}, {Cinabro}, {Filippenko}, {Frieman}, {Galbany}, {Garnavich}, {Goobar}, {Hopp}, {Ihara}, {Jha}, {Kessler}, {Lampeitl}, {Marriner}, {Miquel}, {Moll{\'a}}, {Nichol}, {Nordin}, {Riess}, {Sako}, {Schneider}, {Sollerman}, {Wheeler}, {{\"O}stman}, {Bizyaev}, {Brewington}, {Malanushenko}, {Malanushenko}, {Oravetz}, {Pan}, {Simmons}, \& {Snedden}}]{2010ApJ...713.1026D}
{Dilday}, B., {Smith}, M., {Bassett}, B., {et~al.} 2010, \apj, 713, 1026, \dodoi{10.1088/0004-637X/713/2/1026}

\bibitem[{{Driessen} {et~al.}(2024){Driessen}, {Barr}, {Buckley}, {Caleb}, {Chen}, {Chen}, {Gromadzki}, {Jankowski}, {Kraan-Korteweg}, {Palmerio}, {Rajwade}, {Tremou}, {Kramer}, {Stappers}, {Vergani}, {Woudt}, {Bezuidenhout}, {Malenta}, {Morello}, {Sanidas}, {Surnis}, \& {Fender}}]{2024MNRAS.527.3659D}
{Driessen}, L.~N., {Barr}, E.~D., {Buckley}, D.~A.~H., {et~al.} 2024, \mnras, 527, 3659, \dodoi{10.1093/mnras/stad3329}

\bibitem[{{Fonseca} {et~al.}(2020){Fonseca}, {Andersen}, {Bhardwaj}, {Chawla}, {Good}, {Josephy}, {Kaspi}, {Masui}, {Mckinven}, {Michilli}, {Pleunis}, {Shin}, {Tendulkar}, {Bandura}, {Boyle}, {Brar}, {Cassanelli}, {Cubranic}, {Dobbs}, {Dong}, {Gaensler}, {Hinshaw}, {Landecker}, {Leung}, {Li}, {Lin}, {Mena-Parra}, {Merryfield}, {Naidu}, {Ng}, {Patel}, {Pen}, {Rafiei-Ravandi}, {Rahman}, {Ransom}, {Scholz}, {Smith}, {Stairs}, {Vanderlinde}, {Yadav}, \& {Zwaniga}}]{2020ApJ...891L...6F}
{Fonseca}, E., {Andersen}, B.~C., {Bhardwaj}, M., {et~al.} 2020, \apjl, 891, L6, \dodoi{10.3847/2041-8213/ab7208}

\bibitem[{{Gajjar} {et~al.}(2018){Gajjar}, {Siemion}, {Price}, {Law}, {Michilli}, {Hessels}, {Chatterjee}, {Archibald}, {Bower}, {Brinkman}, {Burke-Spolaor}, {Cordes}, {Croft}, {Enriquez}, {Foster}, {Gizani}, {Hellbourg}, {Isaacson}, {Kaspi}, {Lazio}, {Lebofsky}, {Lynch}, {MacMahon}, {McLaughlin}, {Ransom}, {Scholz}, {Seymour}, {Spitler}, {Tendulkar}, {Werthimer}, \& {Zhang}}]{2018ApJ...863....2G}
{Gajjar}, V., {Siemion}, A.~P.~V., {Price}, D.~C., {et~al.} 2018, \apj, 863, 2, \dodoi{10.3847/1538-4357/aad005}

\bibitem[{{Gehrels}(1986)}]{1986ApJ...303..336G}
{Gehrels}, N. 1986, \apj, 303, 336, \dodoi{10.1086/164079}

\bibitem[{{Glowacki} {et~al.}(2023){Glowacki}, {Lee-Waddell}, {Deller}, {Deg}, {Gordon}, {Grundy}, {Marnoch}, {Shen}, {Ryder}, {Shannon}, {Wong}, {D{\'e}nes}, {Koribalski}, {Murugeshan}, {Rhee}, {Westmeier}, {Bhandari}, {Bosma}, {Holwerda}, \& {Prochaska}}]{2023ApJ...949...25G}
{Glowacki}, M., {Lee-Waddell}, K., {Deller}, A.~T., {et~al.} 2023, \apj, 949, 25, \dodoi{10.3847/1538-4357/acc1e3}

\bibitem[{{Gordon} {et~al.}(2023){Gordon}, {Fong}, {Kilpatrick}, {Eftekhari}, {Leja}, {Prochaska}, {Nugent}, {Bhandari}, {Blanchard}, {Caleb}, {Day}, {Deller}, {Dong}, {Glowacki}, {Gourdji}, {Mannings}, {Mahoney}, {Marnoch}, {Miller}, {Paterson}, {Rastinejad}, {Ryder}, {Sadler}, {Scott}, {Sears}, {Shannon}, {Simha}, {Stappers}, \& {Tejos}}]{2023ApJ...954...80G}
{Gordon}, A.~C., {Fong}, W.-f., {Kilpatrick}, C.~D., {et~al.} 2023, \apj, 954, 80, \dodoi{10.3847/1538-4357/ace5aa}

\bibitem[{{Guo} \& {Wei}(2022)}]{2022JCAP...07..010G}
{Guo}, H.-Y., \& {Wei}, H. 2022, JCAP, 2022, 010, \dodoi{10.1088/1475-7516/2022/07/010}

\bibitem[{{Hashimoto} {et~al.}(2022){Hashimoto}, {Goto}, {Chen}, {Ho}, {Hsiao}, {Wong}, {On}, {Kim}, {Kilerci-Eser}, {Huang}, {Santos}, \& {Yamasaki}}]{2022MNRAS.511.1961H}
{Hashimoto}, T., {Goto}, T., {Chen}, B.~H., {et~al.} 2022, \mnras, 511, 1961, \dodoi{10.1093/mnras/stac065}

\bibitem[{{Heintz} {et~al.}(2020){Heintz}, {Prochaska}, {Simha}, {Platts}, {Fong}, {Tejos}, {Ryder}, {Aggerwal}, {Bhandari}, {Day}, {Deller}, {Kilpatrick}, {Law}, {Macquart}, {Mannings}, {Marnoch}, {Sadler}, \& {Shannon}}]{2020ApJ...903..152H}
{Heintz}, K.~E., {Prochaska}, J.~X., {Simha}, S., {et~al.} 2020, \apj, 903, 152, \dodoi{10.3847/1538-4357/abb6fb}

\bibitem[{{Ibik} {et~al.}(2024){Ibik}, {Drout}, {Gaensler}, {Scholz}, {Michilli}, {Bhardwaj}, {Kaspi}, {Pleunis}, {Cassanelli}, {Cook}, {Dong}, {Kaczmarek}, {Leung}, {Lu}, {Masui}, {Pearlman}, {Rafiei-Ravandi}, {Sand}, {Shin}, {Smith}, \& {Stairs}}]{2024ApJ...961...99I}
{Ibik}, A.~L., {Drout}, M.~R., {Gaensler}, B.~M., {et~al.} 2024, \apj, 961, 99, \dodoi{10.3847/1538-4357/ad0893}

\bibitem[{{James} {et~al.}(2022{\natexlab{a}}){James}, {Prochaska}, {Macquart}, {North-Hickey}, {Bannister}, \& {Dunning}}]{2022MNRAS.510L..18J}
{James}, C.~W., {Prochaska}, J.~X., {Macquart}, J.~P., {et~al.} 2022{\natexlab{a}}, \mnras, 510, L18, \dodoi{10.1093/mnrasl/slab117}

\bibitem[{{James} {et~al.}(2022{\natexlab{b}}){James}, {Ghosh}, {Prochaska}, {Bannister}, {Bhandari}, {Day}, {Deller}, {Glowacki}, {Gordon}, {Heintz}, {Marnoch}, {Ryder}, {Scott}, {Shannon}, \& {Tejos}}]{2022MNRAS.516.4862J}
{James}, C.~W., {Ghosh}, E.~M., {Prochaska}, J.~X., {et~al.} 2022{\natexlab{b}}, \mnras, 516, 4862, \dodoi{10.1093/mnras/stac2524}

\bibitem[{{Kaur} {et~al.}(2022){Kaur}, {Kanekar}, \& {Prochaska}}]{2022ApJ...925L..20K}
{Kaur}, B., {Kanekar}, N., \& {Prochaska}, J.~X. 2022, \apjl, 925, L20, \dodoi{10.3847/2041-8213/ac4ca8}

\bibitem[{{Keane} \& {Kramer}(2008)}]{2008MNRAS.391.2009K}
{Keane}, E.~F., \& {Kramer}, M. 2008, \mnras, 391, 2009, \dodoi{10.1111/j.1365-2966.2008.14045.x}

\bibitem[{{Keane} {et~al.}(2016){Keane}, {Johnston}, {Bhandari}, {Barr}, {Bhat}, {Burgay}, {Caleb}, {Flynn}, {Jameson}, {Kramer}, {Petroff}, {Possenti}, {van Straten}, {Bailes}, {Burke-Spolaor}, {Eatough}, {Stappers}, {Totani}, {Honma}, {Furusawa}, {Hattori}, {Morokuma}, {Niino}, {Sugai}, {Terai}, {Tominaga}, {Yamasaki}, {Yasuda}, {Allen}, {Cooke}, {Jencson}, {Kasliwal}, {Kaplan}, {Tingay}, {Williams}, {Wayth}, {Chandra}, {Perrodin}, {Berezina}, {Mickaliger}, \& {Bassa}}]{2016Natur.530..453K}
{Keane}, E.~F., {Johnston}, S., {Bhandari}, S., {et~al.} 2016, \nat, 530, 453, \dodoi{10.1038/nature17140}

\bibitem[{{Kirsten} {et~al.}(2022){Kirsten}, {Marcote}, {Nimmo}, {Hessels}, {Bhardwaj}, {Tendulkar}, {Keimpema}, {Yang}, {Snelders}, {Scholz}, {Pearlman}, {Law}, {Peters}, {Giroletti}, {Paragi}, {Bassa}, {Hewitt}, {Bach}, {Bezrukovs}, {Burgay}, {Buttaccio}, {Conway}, {Corongiu}, {Feiler}, {Forss{\'e}n}, {Gawro{\'n}ski}, {Karuppusamy}, {Kharinov}, {Lindqvist}, {Maccaferri}, {Melnikov}, {Ould-Boukattine}, {Possenti}, {Surcis}, {Wang}, {Yuan}, {Aggarwal}, {Anna-Thomas}, {Bower}, {Blaauw}, {Burke-Spolaor}, {Cassanelli}, {Clarke}, {Fonseca}, {Gaensler}, {Gopinath}, {Kaspi}, {Kassim}, {Lazio}, {Leung}, {Li}, {Lin}, {Masui}, {Mckinven}, {Michilli}, {Mikhailov}, {Ng}, {Orbidans}, {Pen}, {Petroff}, {Rahman}, {Ransom}, {Shin}, {Smith}, {Stairs}, \& {Vlemmings}}]{2022Natur.602..585K}
{Kirsten}, F., {Marcote}, B., {Nimmo}, K., {et~al.} 2022, \nat, 602, 585, \dodoi{10.1038/s41586-021-04354-w}

\bibitem[{{Law} {et~al.}(2020){Law}, {Butler}, {Prochaska}, {Zackay}, {Burke-Spolaor}, {Mannings}, {Tejos}, {Josephy}, {Andersen}, {Chawla}, {Heintz}, {Aggarwal}, {Bower}, {Demorest}, {Kilpatrick}, {Lazio}, {Linford}, {Mckinven}, {Tendulkar}, \& {Simha}}]{2020ApJ...899..161L}
{Law}, C.~J., {Butler}, B.~J., {Prochaska}, J.~X., {et~al.} 2020, \apj, 899, 161, \dodoi{10.3847/1538-4357/aba4ac}

\bibitem[{{Law} {et~al.}(2024){Law}, {Sharma}, {Ravi}, {Chen}, {Catha}, {Connor}, {Faber}, {Hallinan}, {Harnach}, {Hellbourg}, {Hobbs}, {Hodge}, {Hodges}, {Lamb}, {Rasmussen}, {Sherman}, {Shi}, {Simard}, {Squillace}, {Weinreb}, {Woody}, \& {Yurk}}]{2024ApJ...967...29L}
{Law}, C.~J., {Sharma}, K., {Ravi}, V., {et~al.} 2024, \apj, 967, 29, \dodoi{10.3847/1538-4357/ad3736}

\bibitem[{{Li} {et~al.}(2021){Li}, {Lin}, {Xiong}, {Ge}, {Li}, {Li}, {Lu}, {Zhang}, {Tuo}, {Nang}, {Zhang}, {Xiao}, {Chen}, {Song}, {Xu}, {Liu}, {Jia}, {Cao}, {Qu}, {Zhang}, {Gu}, {Liao}, {Zhao}, {Tan}, {Nie}, {Zhao}, {Zheng}, {Zheng}, {Luo}, {Cai}, {Li}, {Xue}, {Bu}, {Chang}, {Chen}, {Chen}, {Chen}, {Chen}, {Chen}, {Cui}, {Cui}, {Deng}, {Dong}, {Du}, {Fu}, {Gao}, {Gao}, {Gao}, {Gu}, {Guan}, {Guo}, {Han}, {Huang}, {Huo}, {Jiang}, {Jiang}, {Jin}, {Jin}, {Kong}, {Li}, {Li}, {Li}, {Li}, {Li}, {Li}, {Li}, {Liang}, {Liu}, {Liu}, {Liu}, {Liu}, {Liu}, {Lu}, {Lu}, {Luo}, {Ma}, {Meng}, {Ou}, {Sai}, {Shang}, {Song}, {Sun}, {Tao}, {Wang}, {Wang}, {Wang}, {Wang}, {Wang}, {Wen}, {Wu}, {Wu}, {Wu}, {Xiao}, {Xu}, {Yang}, {Yang}, {Yang}, {Yang}, {Yi}, {Yin}, {You}, {Zhang}, {Zhang}, {Zhang}, {Zhang}, {Zhang}, {Zhang}, {Zhang}, {Zhang}, {Zhang}, {Zhang}, {Zhang}, {Zhang}, {Zhang}, {Zhang}, {Zhang}, {Zhang}, {Zhou}, {Zhou}, {Zhu}, {Zhu}, \& {Zhuang}}]{2021NatAs...5..378L}
{Li}, C.~K., {Lin}, L., {Xiong}, S.~L., {et~al.} 2021, Nature Astronomy, 5, 378, \dodoi{10.1038/s41550-021-01302-6}

\bibitem[{{Lorimer} {et~al.}(2007){Lorimer}, {Bailes}, {McLaughlin}, {Narkevic}, \& {Crawford}}]{2007Sci...318..777L}
{Lorimer}, D.~R., {Bailes}, M., {McLaughlin}, M.~A., {Narkevic}, D.~J., \& {Crawford}, F. 2007, Science, 318, 777, \dodoi{10.1126/science.1147532}

\bibitem[{{Luo} {et~al.}(2018){Luo}, {Lee}, {Lorimer}, \& {Zhang}}]{2018MNRAS.481.2320L}
{Luo}, R., {Lee}, K., {Lorimer}, D.~R., \& {Zhang}, B. 2018, \mnras, 481, 2320, \dodoi{10.1093/mnras/sty2364}

\bibitem[{{Luo} {et~al.}(2020){Luo}, {Men}, {Lee}, {Wang}, {Lorimer}, \& {Zhang}}]{2020MNRAS.494..665L}
{Luo}, R., {Men}, Y., {Lee}, K., {et~al.} 2020, \mnras, 494, 665, \dodoi{10.1093/mnras/staa704}

\bibitem[{{Macquart} {et~al.}(2020){Macquart}, {Prochaska}, {McQuinn}, {Bannister}, {Bhandari}, {Day}, {Deller}, {Ekers}, {James}, {Marnoch}, {Os{\l}owski}, {Phillips}, {Ryder}, {Scott}, {Shannon}, \& {Tejos}}]{2020Natur.581..391M}
{Macquart}, J.~P., {Prochaska}, J.~X., {McQuinn}, M., {et~al.} 2020, \nat, 581, 391, \dodoi{10.1038/s41586-020-2300-2}

\bibitem[{{Mahony} {et~al.}(2018){Mahony}, {Ekers}, {Macquart}, {Sadler}, {Bannister}, {Bhandari}, {Flynn}, {Koribalski}, {Prochaska}, {Ryder}, {Shannon}, {Tejos}, {Whiting}, \& {Wong}}]{2018ApJ...867L..10M}
{Mahony}, E.~K., {Ekers}, R.~D., {Macquart}, J.-P., {et~al.} 2018, \apjl, 867, L10, \dodoi{10.3847/2041-8213/aae7cb}

\bibitem[{{Mannings} {et~al.}(2021){Mannings}, {Fong}, {Simha}, {Prochaska}, {Rafelski}, {Kilpatrick}, {Tejos}, {Heintz}, {Bannister}, {Bhandari}, {Day}, {Deller}, {Ryder}, {Shannon}, \& {Tendulkar}}]{2021ApJ...917...75M}
{Mannings}, A.~G., {Fong}, W.-f., {Simha}, S., {et~al.} 2021, \apj, 917, 75, \dodoi{10.3847/1538-4357/abff56}

\bibitem[{{Marcote} {et~al.}(2020){Marcote}, {Nimmo}, {Hessels}, {Tendulkar}, {Bassa}, {Paragi}, {Keimpema}, {Bhardwaj}, {Karuppusamy}, {Kaspi}, {Law}, {Michilli}, {Aggarwal}, {Andersen}, {Archibald}, {Bandura}, {Bower}, {Boyle}, {Brar}, {Burke-Spolaor}, {Butler}, {Cassanelli}, {Chawla}, {Demorest}, {Dobbs}, {Fonseca}, {Giri}, {Good}, {Gourdji}, {Josephy}, {Kirichenko}, {Kirsten}, {Landecker}, {Lang}, {Lazio}, {Li}, {Lin}, {Linford}, {Masui}, {Mena-Parra}, {Naidu}, {Ng}, {Patel}, {Pen}, {Pleunis}, {Rafiei-Ravandi}, {Rahman}, {Renard}, {Scholz}, {Siegel}, {Smith}, {Stairs}, {Vanderlinde}, \& {Zwaniga}}]{2020Natur.577..190M}
{Marcote}, B., {Nimmo}, K., {Hessels}, J.~W.~T., {et~al.} 2020, \nat, 577, 190, \dodoi{10.1038/s41586-019-1866-z}

\bibitem[{{Margalit} {et~al.}(2019){Margalit}, {Berger}, \& {Metzger}}]{2019ApJ...886..110M}
{Margalit}, B., {Berger}, E., \& {Metzger}, B.~D. 2019, \apj, 886, 110, \dodoi{10.3847/1538-4357/ab4c31}

\bibitem[{{Mereghetti} {et~al.}(2020){Mereghetti}, {Savchenko}, {Ferrigno}, {G{\"o}tz}, {Rigoselli}, {Tiengo}, {Bazzano}, {Bozzo}, {Coleiro}, {Courvoisier}, {Doyle}, {Goldwurm}, {Hanlon}, {Jourdain}, {von Kienlin}, {Lutovinov}, {Martin-Carrillo}, {Molkov}, {Natalucci}, {Onori}, {Panessa}, {Rodi}, {Rodriguez}, {S{\'a}nchez-Fern{\'a}ndez}, {Sunyaev}, \& {Ubertini}}]{2020ApJ...898L..29M}
{Mereghetti}, S., {Savchenko}, V., {Ferrigno}, C., {et~al.} 2020, \apjl, 898, L29, \dodoi{10.3847/2041-8213/aba2cf}

\bibitem[{{Metzger} {et~al.}(2017){Metzger}, {Berger}, \& {Margalit}}]{2017ApJ...841...14M}
{Metzger}, B.~D., {Berger}, E., \& {Margalit}, B. 2017, \apj, 841, 14, \dodoi{10.3847/1538-4357/aa633d}

\bibitem[{{Michilli} {et~al.}(2023){Michilli}, {Bhardwaj}, {Brar}, {Gaensler}, {Kaspi}, {Kirichenko}, {Masui}, {Mckinven}, {Ng}, {Patel}, {Sand}, {Scholz}, {Shin}, {Siegel}, {Stairs}, {Cassanelli}, {Cook}, {Dobbs}, {Dong}, {Fonseca}, {Ibik}, {Kaczmarek}, {Leung}, {Pearlman}, {Petroff}, {Pleunis}, {Rafiei-Ravandi}, {Sanghavi}, {Shaw}, \& {Tendulkar}}]{2023ApJ...950..134M}
{Michilli}, D., {Bhardwaj}, M., {Brar}, C., {et~al.} 2023, \apj, 950, 134, \dodoi{10.3847/1538-4357/accf89}

\bibitem[{{Nicholl} {et~al.}(2017){Nicholl}, {Williams}, {Berger}, {Villar}, {Alexander}, {Eftekhari}, \& {Metzger}}]{2017ApJ...843...84N}
{Nicholl}, M., {Williams}, P.~K.~G., {Berger}, E., {et~al.} 2017, \apj, 843, 84, \dodoi{10.3847/1538-4357/aa794d}

\bibitem[{{Niu} {et~al.}(2022){Niu}, {Aggarwal}, {Li}, {Zhang}, {Chatterjee}, {Tsai}, {Yu}, {Law}, {Burke-Spolaor}, {Cordes}, {Zhang}, {Ocker}, {Yao}, {Wang}, {Feng}, {Niino}, {Bochenek}, {Cruces}, {Connor}, {Jiang}, {Dai}, {Luo}, {Li}, {Miao}, {Niu}, {Anna-Thomas}, {Sydnor}, {Stern}, {Wang}, {Yuan}, {Yue}, {Zhou}, {Yan}, {Zhu}, \& {Zhang}}]{2022Natur.606..873N}
{Niu}, C.~H., {Aggarwal}, K., {Li}, D., {et~al.} 2022, \nat, 606, 873, \dodoi{10.1038/s41586-022-04755-5}

\bibitem[{{Ofek}(2007)}]{2007ApJ...659..339O}
{Ofek}, E.~O. 2007, \apj, 659, 339, \dodoi{10.1086/511147}

\bibitem[{{Pastor-Marazuela} {et~al.}(2021){Pastor-Marazuela}, {Connor}, {van Leeuwen}, {Maan}, {ter Veen}, {Bilous}, {Oostrum}, {Petroff}, {Straal}, {Vohl}, {Attema}, {Boersma}, {Kooistra}, {van der Schuur}, {Sclocco}, {Smits}, {Adams}, {Adebahr}, {de Blok}, {Coolen}, {Damstra}, {D{\'e}nes}, {Hess}, {van der Hulst}, {Hut}, {Ivashina}, {Kutkin}, {Loose}, {Lucero}, {Mika}, {Moss}, {Mulder}, {Norden}, {Oosterloo}, {Orr{\'u}}, {Ruiter}, \& {Wijnholds}}]{2021Natur.596..505P}
{Pastor-Marazuela}, I., {Connor}, L., {van Leeuwen}, J., {et~al.} 2021, \nat, 596, 505, \dodoi{10.1038/s41586-021-03724-8}

\bibitem[{{Petroff} {et~al.}(2019){Petroff}, {Hessels}, \& {Lorimer}}]{2019A&ARv..27....4P}
{Petroff}, E., {Hessels}, J.~W.~T., \& {Lorimer}, D.~R. 2019, \aapr, 27, 4, \dodoi{10.1007/s00159-019-0116-6}

\bibitem[{{Petroff} {et~al.}(2022){Petroff}, {Hessels}, \& {Lorimer}}]{2022A&ARv..30....2P}
---. 2022, \aapr, 30, 2, \dodoi{10.1007/s00159-022-00139-w}

\bibitem[{{Piro} \& {Gaensler}(2018)}]{2018ApJ...861..150P}
{Piro}, A.~L., \& {Gaensler}, B.~M. 2018, \apj, 861, 150, \dodoi{10.3847/1538-4357/aac9bc}

\bibitem[{{Planck Collaboration} {et~al.}(2020){Planck Collaboration}, {Aghanim}, {Akrami}, {Ashdown}, {Aumont}, {Baccigalupi}, {Ballardini}, {Banday}, {Barreiro}, {Bartolo}, {Basak}, {Battye}, {Benabed}, {Bernard}, {Bersanelli}, {Bielewicz}, {Bock}, {Bond}, {Borrill}, {Bouchet}, {Boulanger}, {Bucher}, {Burigana}, {Butler}, {Calabrese}, {Cardoso}, {Carron}, {Challinor}, {Chiang}, {Chluba}, {Colombo}, {Combet}, {Contreras}, {Crill}, {Cuttaia}, {de Bernardis}, {de Zotti}, {Delabrouille}, {Delouis}, {Di Valentino}, {Diego}, {Dor{\'e}}, {Douspis}, {Ducout}, {Dupac}, {Dusini}, {Efstathiou}, {Elsner}, {En{\ss}lin}, {Eriksen}, {Fantaye}, {Farhang}, {Fergusson}, {Fernandez-Cobos}, {Finelli}, {Forastieri}, {Frailis}, {Fraisse}, {Franceschi}, {Frolov}, {Galeotta}, {Galli}, {Ganga}, {G{\'e}nova-Santos}, {Gerbino}, {Ghosh}, {Gonz{\'a}lez-Nuevo}, {G{\'o}rski}, {Gratton}, {Gruppuso}, {Gudmundsson}, {Hamann}, {Handley}, {Hansen}, {Herranz}, {Hildebrandt}, {Hivon}, {Huang}, {Jaffe}, {Jones}, {Karakci}, {Keih{\"a}nen},
  {Keskitalo}, {Kiiveri}, {Kim}, {Kisner}, {Knox}, {Krachmalnicoff}, {Kunz}, {Kurki-Suonio}, {Lagache}, {Lamarre}, {Lasenby}, {Lattanzi}, {Lawrence}, {Le Jeune}, {Lemos}, {Lesgourgues}, {Levrier}, {Lewis}, {Liguori}, {Lilje}, {Lilley}, {Lindholm}, {L{\'o}pez-Caniego}, {Lubin}, {Ma}, {Mac{\'\i}as-P{\'e}rez}, {Maggio}, {Maino}, {Mandolesi}, {Mangilli}, {Marcos-Caballero}, {Maris}, {Martin}, {Martinelli}, {Mart{\'\i}nez-Gonz{\'a}lez}, {Matarrese}, {Mauri}, {McEwen}, {Meinhold}, {Melchiorri}, {Mennella}, {Migliaccio}, {Millea}, {Mitra}, {Miville-Desch{\^e}nes}, {Molinari}, {Montier}, {Morgante}, {Moss}, {Natoli}, {N{\o}rgaard-Nielsen}, {Pagano}, {Paoletti}, {Partridge}, {Patanchon}, {Peiris}, {Perrotta}, {Pettorino}, {Piacentini}, {Polastri}, {Polenta}, {Puget}, {Rachen}, {Reinecke}, {Remazeilles}, {Renzi}, {Rocha}, {Rosset}, {Roudier}, {Rubi{\~n}o-Mart{\'\i}n}, {Ruiz-Granados}, {Salvati}, {Sandri}, {Savelainen}, {Scott}, {Shellard}, {Sirignano}, {Sirri}, {Spencer}, {Sunyaev}, {Suur-Uski}, {Tauber}, {Tavagnacco},
  {Tenti}, {Toffolatti}, {Tomasi}, {Trombetti}, {Valenziano}, {Valiviita}, {Van Tent}, {Vibert}, {Vielva}, {Villa}, {Vittorio}, {Wandelt}, {Wehus}, {White}, {White}, {Zacchei}, \& {Zonca}}]{2020A&A...641A...6P}
{Planck Collaboration}, {Aghanim}, N., {Akrami}, Y., {et~al.} 2020, \aap, 641, A6, \dodoi{10.1051/0004-6361/201833910}

\bibitem[{{Price} {et~al.}(2021){Price}, {Flynn}, \& {Deller}}]{2021PASA...38...38P}
{Price}, D.~C., {Flynn}, C., \& {Deller}, A. 2021, \pasa, 38, e038, \dodoi{10.1017/pasa.2021.33}

\bibitem[{{Prochaska} {et~al.}(2019){Prochaska}, {Macquart}, {McQuinn}, {Simha}, {Shannon}, {Day}, {Marnoch}, {Ryder}, {Deller}, {Bannister}, {Bhandari}, {Bordoloi}, {Bunton}, {Cho}, {Flynn}, {Mahony}, {Phillips}, {Qiu}, \& {Tejos}}]{2019Sci...366..231P}
{Prochaska}, J.~X., {Macquart}, J.-P., {McQuinn}, M., {et~al.} 2019, Science, 366, 231, \dodoi{10.1126/science.aay0073}

\bibitem[{{Rajwade} {et~al.}(2020){Rajwade}, {Mickaliger}, {Stappers}, {Morello}, {Agarwal}, {Bassa}, {Breton}, {Caleb}, {Karastergiou}, {Keane}, \& {Lorimer}}]{2020MNRAS.495.3551R}
{Rajwade}, K.~M., {Mickaliger}, M.~B., {Stappers}, B.~W., {et~al.} 2020, \mnras, 495, 3551, \dodoi{10.1093/mnras/staa1237}

\bibitem[{{Ravi}(2019)}]{2019NatAs...3..928R}
{Ravi}, V. 2019, Nature Astronomy, 3, 928, \dodoi{10.1038/s41550-019-0831-y}

\bibitem[{{Ravi} {et~al.}(2019){Ravi}, {Catha}, {D'Addario}, {Djorgovski}, {Hallinan}, {Hobbs}, {Kocz}, {Kulkarni}, {Shi}, {Vedantham}, {Weinreb}, \& {Woody}}]{2019Natur.572..352R}
{Ravi}, V., {Catha}, M., {D'Addario}, L., {et~al.} 2019, \nat, 572, 352, \dodoi{10.1038/s41586-019-1389-7}

\bibitem[{{Ravi} {et~al.}(2022){Ravi}, {Law}, {Li}, {Aggarwal}, {Bhardwaj}, {Burke-Spolaor}, {Connor}, {Lazio}, {Simard}, {Somalwar}, \& {Tendulkar}}]{2022MNRAS.513..982R}
{Ravi}, V., {Law}, C.~J., {Li}, D., {et~al.} 2022, \mnras, 513, 982, \dodoi{10.1093/mnras/stac465}

\bibitem[{{Ravi} {et~al.}(2023{\natexlab{a}}){Ravi}, {Catha}, {Chen}, {Connor}, {Cordes}, {Faber}, {Lamb}, {Hallinan}, {Harnach}, {Hellbourg}, {Hobbs}, {Hodge}, {Hodges}, {Law}, {Rasmussen}, {Sharma}, {Sherman}, {Shi}, {Simard}, {Somalwar}, {Squillace}, {Weinreb}, {Woody}, \& {Yadlapalli}}]{2023arXiv230101000R}
{Ravi}, V., {Catha}, M., {Chen}, G., {et~al.} 2023{\natexlab{a}}, arXiv e-prints, arXiv:2301.01000, \dodoi{10.48550/arXiv.2301.01000}

\bibitem[{{Ravi} {et~al.}(2023{\natexlab{b}}){Ravi}, {Catha}, {Chen}, {Connor}, {Faber}, {Lamb}, {Hallinan}, {Harnach}, {Hellbourg}, {Hobbs}, {Hodge}, {Hodges}, {Law}, {Rasmussen}, {Sharma}, {Sherman}, {Shi}, {Simard}, {Squillace}, {Weinreb}, {Woody}, {Yadlapalli}, {Ahumada}, {Dong}, {Fremling}, {Huang}, {Karambelkar}, \& {Miller}}]{2023ApJ...949L...3R}
---. 2023{\natexlab{b}}, \apjl, 949, L3, \dodoi{10.3847/2041-8213/acc4b6}

\bibitem[{{Ridnaia} {et~al.}(2021){Ridnaia}, {Svinkin}, {Frederiks}, {Bykov}, {Popov}, {Aptekar}, {Golenetskii}, {Lysenko}, {Tsvetkova}, {Ulanov}, \& {Cline}}]{2021NatAs...5..372R}
{Ridnaia}, A., {Svinkin}, D., {Frederiks}, D., {et~al.} 2021, Nature Astronomy, 5, 372, \dodoi{10.1038/s41550-020-01265-0}

\bibitem[{{Ryder} {et~al.}(2023){Ryder}, {Bannister}, {Bhandari}, {Deller}, {Ekers}, {Glowacki}, {Gordon}, {Gourdji}, {James}, {Kilpatrick}, {Lu}, {Marnoch}, {Moss}, {Prochaska}, {Qiu}, {Sadler}, {Simha}, {Sammons}, {Scott}, {Tejos}, \& {Shannon}}]{2023Sci...382..294R}
{Ryder}, S.~D., {Bannister}, K.~W., {Bhandari}, S., {et~al.} 2023, Science, 382, 294, \dodoi{10.1126/science.adf2678}

\bibitem[{{Sharma} {et~al.}(2023){Sharma}, {Somalwar}, {Law}, {Ravi}, {Catha}, {Chen}, {Connor}, {Faber}, {Hallinan}, {Harnach}, {Hellbourg}, {Hobbs}, {Hodge}, {Hodges}, {Lamb}, {Rasmussen}, {Sherman}, {Shi}, {Simard}, {Squillace}, {Weinreb}, {Woody}, {Yadlapalli}, \& {Deep Synoptic Array Team}}]{2023ApJ...950..175S}
{Sharma}, K., {Somalwar}, J., {Law}, C., {et~al.} 2023, \apj, 950, 175, \dodoi{10.3847/1538-4357/accf1d}

\bibitem[{{Shin} {et~al.}(2023){Shin}, {Masui}, {Bhardwaj}, {Cassanelli}, {Chawla}, {Dobbs}, {Dong}, {Fonseca}, {Gaensler}, {Herrera-Mart{\'\i}n}, {Kaczmarek}, {Kaspi}, {Leung}, {Merryfield}, {Michilli}, {M{\"u}nchmeyer}, {Pearlman}, {Rafiei-Ravandi}, {Smith}, {Stairs}, \& {Tendulkar}}]{2023ApJ...944..105S}
{Shin}, K., {Masui}, K.~W., {Bhardwaj}, M., {et~al.} 2023, \apj, 944, 105, \dodoi{10.3847/1538-4357/acaf06}

\bibitem[{{Shull} \& {Danforth}(2018)}]{2018ApJ...852L..11S}
{Shull}, J.~M., \& {Danforth}, C.~W. 2018, \apjl, 852, L11, \dodoi{10.3847/2041-8213/aaa2fa}

\bibitem[{{Shull} {et~al.}(2012){Shull}, {Smith}, \& {Danforth}}]{2012ApJ...759...23S}
{Shull}, J.~M., {Smith}, B.~D., \& {Danforth}, C.~W. 2012, \apj, 759, 23, \dodoi{10.1088/0004-637X/759/1/23}

\bibitem[{{Spitler} {et~al.}(2014){Spitler}, {Cordes}, {Hessels}, {Lorimer}, {McLaughlin}, {Chatterjee}, {Crawford}, {Deneva}, {Kaspi}, {Wharton}, {Allen}, {Bogdanov}, {Brazier}, {Camilo}, {Freire}, {Jenet}, {Karako-Argaman}, {Knispel}, {Lazarus}, {Lee}, {van Leeuwen}, {Lynch}, {Ransom}, {Scholz}, {Siemens}, {Stairs}, {Stovall}, {Swiggum}, {Venkataraman}, {Zhu}, {Aulbert}, \& {Fehrmann}}]{2014ApJ...790..101S}
{Spitler}, L.~G., {Cordes}, J.~M., {Hessels}, J.~W.~T., {et~al.} 2014, \apj, 790, 101, \dodoi{10.1088/0004-637X/790/2/101}

\bibitem[{{Spitler} {et~al.}(2016){Spitler}, {Scholz}, {Hessels}, {Bogdanov}, {Brazier}, {Camilo}, {Chatterjee}, {Cordes}, {Crawford}, {Deneva}, {Ferdman}, {Freire}, {Kaspi}, {Lazarus}, {Lynch}, {Madsen}, {McLaughlin}, {Patel}, {Ransom}, {Seymour}, {Stairs}, {Stappers}, {van Leeuwen}, \& {Zhu}}]{2016Natur.531..202S}
{Spitler}, L.~G., {Scholz}, P., {Hessels}, J.~W.~T., {et~al.} 2016, \nat, 531, 202, \dodoi{10.1038/nature17168}

\bibitem[{{Sridhar} {et~al.}(2021){Sridhar}, {Metzger}, {Beniamini}, {Margalit}, {Renzo}, {Sironi}, \& {Kovlakas}}]{2021ApJ...917...13S}
{Sridhar}, N., {Metzger}, B.~D., {Beniamini}, P., {et~al.} 2021, \apj, 917, 13, \dodoi{10.3847/1538-4357/ac0140}

\bibitem[{{Tavani} {et~al.}(2021){Tavani}, {Casentini}, {Ursi}, {Verrecchia}, {Addis}, {Antonelli}, {Argan}, {Barbiellini}, {Baroncelli}, {Bernardi}, {Bianchi}, {Bulgarelli}, {Caraveo}, {Cardillo}, {Cattaneo}, {Chen}, {Costa}, {Del Monte}, {Di Cocco}, {Di Persio}, {Donnarumma}, {Evangelista}, {Feroci}, {Ferrari}, {Fioretti}, {Fuschino}, {Galli}, {Gianotti}, {Giuliani}, {Labanti}, {Lazzarotto}, {Lipari}, {Longo}, {Lucarelli}, {Magro}, {Marisaldi}, {Mereghetti}, {Morelli}, {Morselli}, {Naldi}, {Pacciani}, {Parmiggiani}, {Paoletti}, {Pellizzoni}, {Perri}, {Perotti}, {Piano}, {Picozza}, {Pilia}, {Pittori}, {Puccetti}, {Pupillo}, {Rapisarda}, {Rappoldi}, {Rubini}, {Setti}, {Soffitta}, {Trifoglio}, {Trois}, {Vercellone}, {Vittorini}, {Giommi}, \& {D'Amico}}]{2021NatAs...5..401T}
{Tavani}, M., {Casentini}, C., {Ursi}, A., {et~al.} 2021, Nature Astronomy, 5, 401, \dodoi{10.1038/s41550-020-01276-x}

\bibitem[{{Tendulkar} {et~al.}(2017){Tendulkar}, {Bassa}, {Cordes}, {Bower}, {Law}, {Chatterjee}, {Adams}, {Bogdanov}, {Burke-Spolaor}, {Butler}, {Demorest}, {Hessels}, {Kaspi}, {Lazio}, {Maddox}, {Marcote}, {McLaughlin}, {Paragi}, {Ransom}, {Scholz}, {Seymour}, {Spitler}, {van Langevelde}, \& {Wharton}}]{2017ApJ...834L...7T}
{Tendulkar}, S.~P., {Bassa}, C.~G., {Cordes}, J.~M., {et~al.} 2017, \apjl, 834, L7, \dodoi{10.3847/2041-8213/834/2/L7}

\bibitem[{{Wang} \& {Nitz}(2022)}]{2022ApJ...937...89W}
{Wang}, Y.-F., \& {Nitz}, A.~H. 2022, \apj, 937, 89, \dodoi{10.3847/1538-4357/ac82ae}

\bibitem[{{Wen} {et~al.}(2016{\natexlab{a}}){Wen}, {Wang}, {Yan}, {Yuan}, {Liu}, {Chen}, \& {Chen}}]{2016Ap&SS.361..261W}
{Wen}, Z.~G., {Wang}, N., {Yan}, W.~M., {et~al.} 2016{\natexlab{a}}, \apss, 361, 261, \dodoi{10.1007/s10509-016-2755-7}

\bibitem[{{Wen} {et~al.}(2016{\natexlab{b}}){Wen}, {Wang}, {Yuan}, {Yan}, {Manchester}, {Yuen}, \& {Gajjar}}]{2016A&A...592A.127W}
{Wen}, Z.~G., {Wang}, N., {Yuan}, J.~P., {et~al.} 2016{\natexlab{b}}, \aap, 592, A127, \dodoi{10.1051/0004-6361/201628214}

\bibitem[{{Wen} {et~al.}(2022){Wen}, {Yuan}, {Wang}, {Li}, {Chen}, {Wang}, {Wu}, {Yan}, {Yuen}, {Wang}, {Tedila}, {Wang}, {Zhu}, {Niu}, {Miao}, {Xue}, {Duan}, {Xiang}, \& {He}}]{2022ApJ...929...71W}
{Wen}, Z.~G., {Yuan}, J.~P., {Wang}, N., {et~al.} 2022, \apj, 929, 71, \dodoi{10.3847/1538-4357/ac5d5d}

\bibitem[{{Xiao} {et~al.}(2021){Xiao}, {Wang}, \& {Dai}}]{2021SCPMA..6449501X}
{Xiao}, D., {Wang}, F., \& {Dai}, Z. 2021, Science China Physics, Mechanics, and Astronomy, 64, 249501, \dodoi{10.1007/s11433-020-1661-7}

\bibitem[{{Xu} {et~al.}(2022){Xu}, {Niu}, {Chen}, {Lee}, {Zhu}, {Dong}, {Zhang}, {Jiang}, {Wang}, {Xu}, {Zhang}, {Fu}, {Filippenko}, {Peng}, {Zhou}, {Zhang}, {Wang}, {Feng}, {Li}, {Brink}, {Li}, {Lu}, {Yang}, {Caballero}, {Cai}, {Chen}, {Dai}, {Djorgovski}, {Esamdin}, {Gan}, {Guhathakurta}, {Han}, {Hao}, {Huang}, {Jiang}, {Li}, {Li}, {Li}, {Li}, {Li}, {Liu}, {Luo}, {Men}, {Niu}, {Peng}, {Qian}, {Song}, {Stern}, {Stockton}, {Sun}, {Wang}, {Wang}, {Wang}, {Wang}, {Wu}, {Xiao}, {Xiong}, {Xu}, {Xu}, {Yang}, {Yang}, {Yao}, {Yi}, {Yue}, {Yu}, {Yu}, {Yuan}, {Zhang}, {Zhang}, {Zhang}, {Zhao}, {Zheng}, {Zhu}, \& {Zou}}]{2022Natur.609..685X}
{Xu}, H., {Niu}, J.~R., {Chen}, P., {et~al.} 2022, \nat, 609, 685, \dodoi{10.1038/s41586-022-05071-8}

\bibitem[{{Xu} {et~al.}(2023){Xu}, {Feng}, {Li}, {Wang}, {Zhang}, {Xie}, {Chen}, {Wang}, {Kang}, {Hu}, {Zheng}, {Tsai}, {Chen}, \& {Zhou}}]{2023Univ....9..330X}
{Xu}, J., {Feng}, Y., {Li}, D., {et~al.} 2023, Universe, 9, 330, \dodoi{10.3390/universe9070330}

\bibitem[{{Yamasaki} \& {Totani}(2020)}]{2020ApJ...888..105Y}
{Yamasaki}, S., \& {Totani}, T. 2020, \apj, 888, 105, \dodoi{10.3847/1538-4357/ab58c4}

\bibitem[{{Yang} \& {Zhang}(2017)}]{2017ApJ...847...22Y}
{Yang}, Y.-P., \& {Zhang}, B. 2017, \apj, 847, 22, \dodoi{10.3847/1538-4357/aa8721}

\bibitem[{{Yao} {et~al.}(2017){Yao}, {Manchester}, \& {Wang}}]{2017ApJ...835...29Y}
{Yao}, J.~M., {Manchester}, R.~N., \& {Wang}, N. 2017, \apj, 835, 29, \dodoi{10.3847/1538-4357/835/1/29}

\bibitem[{{Zhang}(2014)}]{2014ApJ...780L..21Z}
{Zhang}, B. 2014, \apjl, 780, L21, \dodoi{10.1088/2041-8205/780/2/L21}

\bibitem[{{Zhang}(2023)}]{2023RvMP...95c5005Z}
---. 2023, Reviews of Modern Physics, 95, 035005, \dodoi{10.1103/RevModPhys.95.035005}

\bibitem[{{Zhang} {et~al.}(2020{\natexlab{a}}){Zhang}, {Yi}, \& {Wang}}]{2020ApJ...893...44Z}
{Zhang}, G.~Q., {Yi}, S.~X., \& {Wang}, F.~Y. 2020{\natexlab{a}}, \apj, 893, 44, \dodoi{10.3847/1538-4357/ab7c5c}

\bibitem[{{Zhang} {et~al.}(2020{\natexlab{b}}){Zhang}, {Yu}, {He}, \& {Wang}}]{2020ApJ...900..170Z}
{Zhang}, G.~Q., {Yu}, H., {He}, J.~H., \& {Wang}, F.~Y. 2020{\natexlab{b}}, \apj, 900, 170, \dodoi{10.3847/1538-4357/abaa4a}

\bibitem[{{Zhang} \& {Zhang}(2022)}]{2022ApJ...924L..14Z}
{Zhang}, R.~C., \& {Zhang}, B. 2022, \apjl, 924, L14, \dodoi{10.3847/2041-8213/ac46ad}

\bibitem[{{Zhang} {et~al.}(2021){Zhang}, {Yan}, {Li}, {Zhang}, \& {Wang}}]{2021ApJ...906...49Z}
{Zhang}, Z.~J., {Yan}, K., {Li}, C.~M., {Zhang}, G.~Q., \& {Wang}, F.~Y. 2021, \apj, 906, 49, \dodoi{10.3847/1538-4357/abceb9}

\bibitem[{{Zhao} {et~al.}(2021){Zhao}, {Zhang}, {Wang}, {Tu}, \& {Wang}}]{2021ApJ...907..111Z}
{Zhao}, Z.~Y., {Zhang}, G.~Q., {Wang}, Y.~Y., {Tu}, Z.-L., \& {Wang}, F.~Y. 2021, \apj, 907, 111, \dodoi{10.3847/1538-4357/abd321}

\bibitem[{{Zhu} {et~al.}(2023){Zhu}, {Niu}, {Cui}, {Li}, {Feng}, {Tsai}, {Wang}, {Zhang}, {Meng}, \& {Zheng}}]{2023Univ....9..251Z}
{Zhu}, Y., {Niu}, C., {Cui}, X., {et~al.} 2023, Universe, 9, 251, \dodoi{10.3390/universe9060251}

\end{thebibliography}
\bibliographystyle{aasjournal}



\end{document}